\documentclass[useAMS]{mn2e}
\usepackage{graphicx}
\usepackage{amsmath}
\usepackage{amssymb}
\usepackage{color}
\usepackage{url}

\newcommand  \acc     {\ifmmode {\rm km\,s}^{-2} \else km\,s$^{-2}$\fi}

\newcommand  \ergs     {\ifmmode {\rm ergs\,s}^{-1} \else ergs s$^{-1}$\fi}
\newcommand  \ergcms   {\ifmmode {\rm erg~cm}^{-2}\,{\rm s}^{-1}
                        \else erg~cm$^{-2}$\,s$^{-1}$\fi}
\newcommand  \ergcmsA  {\ifmmode{\rm erg\,cm}^{-2}\,{\rm s}^{-1}\,{\rm\AA}^{-1}
                        \else ergs\,cm$^{-2}$\,s$^{-1}$\,\AA$^{-1}$\fi}
\newcommand  \ergcmsHz {\ifmmode{\rm ergs\,cm}^{-2}\,{\rm s}^{-1}\,{\rm Hz}^{-1}
                        \else ergs\,cm$^{-2}$\,s$^{-1}$\,Hz$^{-1}$\fi}
\newcommand  \phcms    {\ifmmode {\rm ph\,cm}^{-2}\,{\rm s}^{-1}
                        \else ph\,cm$^{-2}$\,s$^{-1}$\fi}
\newcommand  \phcmsA   {\ifmmode {\rm ph\,cm}^{-2}\,{\rm s}^{-1}\,{\rm\AA}^{-1}
                        \else ph\,cm$^{-2}$\,s$^{-1}$\,\AA$^{-1}$\fi}

\newcommand\aj{{AJ}}%
\newcommand\araa{{ARA\&A}}%
\newcommand\apj{{ApJ}}%
\newcommand\apjl{{ApJ}}%
\newcommand\apjs{{ApJS}}%
\newcommand\aap{{A\&A}}%
\newcommand\mnras{{MNRAS}}%
\newcommand\pasj{{PASJ}}%
\newcommand\nat{{Nature}}%
\title[LOSS Supernova Delay Time Distribution]
{
Nearby Supernova Rates from the Lick Observatory Supernova Search.
IV. A Recovery Method for the Delay Time Distribution}
\author[D. Maoz et al.]
{Dan Maoz$^{1}$\thanks{E-mail: maoz@astro.tau.ac.il},
Filippo Mannucci$^{2}$, Weidong Li$^{3}$, Alexei V. Filippenko$^{3}$,
\newauthor
Massimo Della Valle$^{4}$, Nino Panagia$^{5,6,7}$ \\
$^{1}$School of Physics and Astronomy, 
Tel-Aviv University, Tel-Aviv 69978,
Israel\\
$^{2}$INAF-Istituto di Radioastronomia, Largo Enrico Fermi 5, Firenze
50125, Italy\\
$^{3}$Department of Astronomy, University of California, Berkeley, CA 94720-3411, USA\\
$^{4}$INAF-Osservatorio Astronomico di Capodimonte, Salita Moiariello
16, Napoli 80131, Italy\\
$^{5}$Space Telescope Science Institute, 3700 San Martin Drive,
Baltimore, MD 21218, USA\\
$^{6}$INAF - Osservatorio Astrofisico di Catania, Via Santa Sofia 78, I-95123 
Catania, Italy\\
$^{7}$Supernova Ltd., OYV \#131, Northsound Road, Virgin Gorda, British Virgin 
Islands\\
} \date{\today}

\begin{document}

\maketitle

\label{firstpage}

\begin{abstract}
Recovery of the hypothetical supernova (SN) delay-time 
distribution (DTD) -- the SN rate versus time
that would follow a brief burst of star formation -- can shed
light on SN progenitors and physics, as well as on the timescales
of chemical enrichment. Previous attempts to reconstruct the DTD have
been based either on comparison of mean SN rates versus redshift to cosmic
star-formation history (SFH), or on the comparison of SN rates among
galaxies with different mean ages. Here, 
we present an approach to recover the SN DTD that avoids the
averaging and loss of information of other schemes. We 
compare the SFHs of {\it individual} 
galaxies to the numbers
of  SNe discovered by a survey in each galaxy (generally zero,
sometimes one SN, rarely a few). 
 We apply the method to a subsample of
3505 galaxies, hosting 82 type-Ia SNe (SNe~Ia) and 119 core-collapse
supernovae (CC~SNe),
 from the Lick Observatory SN Search (LOSS), that have SFHs
 reconstructed from Sloan Digital Sky Survey (SDSS) spectra. 
We find a $> 2\sigma$  SN~Ia DTD 
signal in our shortest-delay, ``prompt'' bin 
at $<420$~Myr. 
We identify and study a systematic error, due to the limited aperture
of the SDSS spectroscopic fibres, that causes some of the prompt signal
to leak to the later bins of the DTD.
Taking this systematic error into account,
 we demonstrate that a prompt SN~Ia contribution is required by
the data at the $>99\%$ confidence level. We further find
a $4\sigma$ indication of
 SNe~Ia that are ``delayed'' by $>2.4$~Gyr. Thus, the data support the
 existence of
 both prompt and delayed SNe~Ia. 
We measure the time integral over the SN DTD. For CC~SNe we find 
a total yield of $0.010\pm0.002$~SNe per M$_\odot$ formed, 
in excellent agreement
with expectations, if all stars more massive than 8 M$_\odot$ lead to
visible SN explosions. This argues against
 scenarios in which the minimum mass for
core-collapse SNe is $\gtrsim 10$ M$_\odot$, or in which a significant fraction
of massive stars collapse without an accompanying explosion. 
For SNe~Ia, the time-integrated yield is $0.0023\pm0.0006$~SNe 
per M$_\odot$ formed, most of them with delays $<2.4$~Gyr. 
Finally, we show the robust performance of the method on
simulated samples, and demonstrate that  its
application to already-existing 
SN samples, such as the full LOSS 
sample, but with complete and unbiased SFH estimates for the survey galaxies,
could provide an accurate and detailed measurement of the SN~Ia DTD.

\end{abstract}

\begin{keywords}
supernovae: general -- methods: data analysis -- galaxies: star formation 
\end{keywords}



\newpage

\section{Introduction}
Supernovae (SNe) figure prominently in many fields, whether
in their roles as calibratable candles for cosmology, as the major sources of 
intermediate-mass and heavy elements, as heaters of the interstellar
medium, as accelerators of cosmic rays, and more. Physically, they are
separated mainly 
into core-collapse SNe (CC~SNe), which occur when the iron core of a 
massive star collapses to form a neutron star or a black hole, and type-Ia
SNe (SNe~Ia), which explode when a degenerate carbon-oxygen stellar
core, probably a white dwarf (WD), approaches (or, rarely, exceeds) the Chandrasekhar
mass, igniting the carbon and triggering a thermonuclear runaway. 
The two paths most often hypothesised
for this mass growth in SNe~Ia are the single-degenerate (SD) scenario,
whereby a WD in a semidetached binary accretes matter from a
main-sequence or evolved normal companion star (Whelan \& Iben 1973), 
and the double-degenerate
(DD) scenario, in which two WDs merge (Iben \& Tutukov 1984; Webbink 1984). 
Additional, less conventional,
paths have also been considered (e.g., Tout 2005; Maoz \& Mannucci 2008; 
Raskin et al. 2009a; Rosswog et al. 2009). 

For CC~SNe, while many questions remain
regarding the progenitors and the physics of particular subtypes,
massive progenitor stars have
by now been identified in pre-explosion images in a growing number of
cases (see Smartt 2009, for a recent summary). This contrasts
with the situation for SNe~Ia, where only one or two very ambiguous
progenitor identifications exist (Voss \& Nelemans 2008;
Roelofs et al. 2008; Gonz{\'a}lez Hern{\'a}ndez  et
al. 2009; Kerzendorf et al. 2009). 
We thus do not quite know what is exploding in a SN~Ia, an
unsatisfactory situation given the ubiquity and importance of these events. 

Driven by these problems,
a major objective of SN studies has been the recovery of the SN
delay-time distribution (DTD). The DTD is the SN rate as a function of
time that would be observed following a $\delta$-function burst of
star formation. (In other contexts, the DTD would be called the delay
function, the transfer function, or the Green's function.) Knowledge of
the DTD would be useful for understanding the route along which 
cosmic metal enrichment and energy input by SNe proceed, 
but no less important,
for obtaining clues about the SN progenitor systems. Different
progenitor stars, binary systems, and binary-evolution scenarios
predict different DTDs. 

The lifetime of a star with the minimum 
initial mass that is thought to lead to 
a CC~SN explosion, $\ga 8$
M$_\odot$, sets a time division between CC~SNe and SNe~Ia in the DTD, 
at $\sim 40$~Myr. 
The precise mass border between core collapse and WD
formation also depends on metallicity. Furthermore,
initially lower-mass stars in tight binaries can become
``rejuvenated'' by mass transfer and explode as CC~SNe somewhat later
than this.

For SNe~Ia, the situation is much less clear. In both of the
currently popular progenitor scenarios, SD and DD, calculations of 
the DTD depend
on a series of assumptions regarding initial conditions (initial mass function, binarity
fraction, mass-ratio distribution, separation distribution), and
complex physics (mass loss, mass transfer, common-envelope
evolution, accretion) that is sometimes computationally 
intractable except in the most
rudimentary, parametrized forms (e.g., Yungelson \& Livio 2000; Hurley et al. 2002; Han \&
Podsiadlowski 2004; Nelemans et al. 2005; Ruiter et al. 2009; 
Bogomazov \& Tutukov 2009; Meng \& Yang 2010; Mennekens et al. 2010).
In principle, observational
estimates of the DTD could rule out particular theoretical models.
Given the theoretical uncertainties, it is probably more
realistic that the observations simply provide a ground truth
 that successful models will need to reproduce.

Previous attempts to recover the DTD have used SN rates  measured 
in surveys of galaxies at different redshifts (i.e., different cosmic
times), compared to cosmic star-formation histories (SFHs). This has been
attempted for field surveys (Gal-Yam \& Maoz 2004; Strolger et
al. 2004; Poznanski et al. 2007; Dahlen et al. 2008) and 
galaxy-cluster surveys (Maoz \& Gal-Yam 2004; Maoz et al. 2010).
An alternative approach has been to look at the SN rates per unit
stellar mass in galaxies
of particular types (star forming, quiescent, etc.), and to attempt
to assign to each type a ``formation age,'' or some generic,
simple, SFH (e.g., Mannucci et al. 2005, 2006;
Sullivan et al. 2006; Totani et al. 2008; Pritchet et al. 2008;
Raskin et al. 2009b).
 
Results have been controversial and often contradictory. For example,
Dahlen et al. (2004, 2008) have argued for a SN~Ia DTD that is peaked
at a delay of $\sim 3$~Gyr, with few SNe~Ia at delays that are much 
shorter or longer. In contrast, Mannucci et al. (2005, 2006),
Scannapieco \& Bildsten (2005), and Sullivan et al. (2006) have found
evidence for the existence of comparable numbers of 
both ``prompt'' and ``delayed'' SNe~Ia:
 the former explode
within $\sim 500$~Myr (or perhaps even within 100~Myr) of star
formation\footnote{We note that diverse delay ranges have been
  associated in the literature with the term ``prompt'' --- e.g.,
  $<100$~Myr (Mannucci et al. 2006); $<180$~Myr (Aubourg et al. 2008);
  200--500~Myr (Raskin et al. 2009b); $<350$, $<700$~Myr, 
or $<1$~Gyr (Scannapieco \&
  Bildsten 2005). The term therefore generally labels delays of roughly
a few hundred Myr. In our analysis of the 
Lick Observatory SN Search herein, we will
define prompt SNe~Ia as those with delays $<420$~Myr.},
while the latter 
may have delays as long as 10~Gyr. The 
 SN~Ia rate has
been described as the sum of two components, one 
proportional to stellar mass and the other proportional to the CC~SN rate
(Mannucci et al. 2005).
In the similar ``$A+B$'' parameterisation
introduced by Scannapieco \& Bildsten (2005), the prompt-component 
rate is proportional to the star-formation rate (SFR). The two SN~Ia
components need not represent two 
distinct physical populations. Instead, they could constitute the 
SNe included in two coarsely sampled time bins of what is in reality 
a continuous DTD.  For example, Pritchet et al. (2008) have argued 
that a  $t^{-0.5\pm 0.2}$ power-law DTD provides an improved fit,
compared to the $A+B$ model, to the dependence of SN rates on galaxy
SFR, as  measured in the Supernova Legacy Survey.
We also note that a truly bimodal DTD, if it exists,
 could arise either from two different
coexisting  progenitor paths (e.g., DD and SD), or from bimodality in
some secondary parameter, such as the binary separation distribution,
 of a single explosion mechanism. Regardless, there is currently 
an unclear picture 
on the form of the DTD, even at the most coarse resolution level.
   
A shortcoming of the approaches described above for recovering the DTD
is that they involve averaging over the galaxy population
(i.e., all the SNe are assumed to come from the entire host population 
considered), or averaging over time (i.e., the detailed SFH of a galaxy is 
represented by a single ``age'' or simplified history for all galaxies
of a certain type). Consequently, 
these approaches involve loss of information,
and potential systematic errors (e.g., due to unrepresentative
simplified histories).  

In this paper, the fourth in a series analysing SN rates from
the Lick Observatory SN Search (LOSS), 
we introduce a new approach to recover the DTD,
by posing DTD inversion as a discretised linear problem.
In this process, we use all
of the available information on the SFHs of
individual galaxies, rather than averaging rates over many types of galaxies  
in some redshift interval, or assigning a mean star-formation-weighted age 
to each galaxy. In \S2 below, we present the method, and in
\S3, we apply the method
to a subsample of LOSS galaxies and their SNe. 
We test the method's performance on simulated SN surveys in \S4.
Our results are summarised and we
discuss some future prospects in \S5.

\section{Reconstruction of the SN DTD --- Method}

Consider a sample of $N$ galaxies.
The SN rate in the $i$th galaxy observed at cosmic 
time $t$ is given by the convolution
\begin{equation}
\label{convolution}
r_i(t)=\int_0^{t} S_i(t-\tau)\Psi (\tau) d\tau ,
\end{equation}
where $S_i(t)$ is the SFR versus cosmic time of the
$i$th galaxy (stellar mass formed per unit
time), $\Psi (\tau)$ is the DTD (SNe per unit time per unit stellar mass
formed), and the integration is from the Big Bang ($t=0$) to the time
of observation. For the purpose of this paper, we assume that the
DTD is a universal function: it is the same in all galaxies,
independent of environment,
metallicity, and cosmic time --- a simplifying assumption that may be
invalid at some level. For example, a dependence of SN delay time 
on metallicity is 
expected in some models (e.g., Kobayashi et al. 2000). Similarly,
variations in the initial mass function (IMF) with
cosmic times or environment would also lead to a variable DTD,
but we will again ignore this possibility in the present context.

In contrast to the averaging approaches followed in the past (see \S1), we will
attempt to recover the DTD by directly inverting
a linear, discretized version of Eq.~\ref{convolution}, where the
detailed history of every individual galaxy or galaxy subunit is taken 
into account. Suppose the SFHs of the
$i=1,2,...,N$ galaxies 
monitored as part of 
a SN survey are known (e.g., based on reconstruction of their stellar 
populations), with a temporal resolution that permits binning the 
stellar mass formed in each galaxy 
into $j=1,2,..,K$ discrete time bins, where 
increasing $j$ corresponds to increasing lookback time. 
The time bins need not necessarily be equal, and generally will not
be, since the temporal resolution of the SFH
reconstruction degrades with increasing lookback time.
For the $i$th galaxy
in the survey, the stellar mass formed in the $j$th time bin is
$m_{ij}$. 
The mean of the DTD over the $j$th bin (corresponding to a
delay range equal to the lookback-time range of the $j$th bin in the
SFH) is $\Psi_j$. Then  the integration in Eq.~\ref{convolution} can
be approximated as a sum,
\begin{equation}
\label{discreteconvolution}
r_i\approx\sum_{j=1}^K m_{ij}\Psi_j  ,
\end{equation}
where $r_i$, the SN rate in a given galaxy, is measured at a
particular cosmic time (e.g., corresponding to the redshift of the
particular SN survey). Given a survey of $N$ galaxies, each with 
an observed SN rate, $r_i$, and a known binned SFH,
$m_{ij}$, one could, in principle, algebraically 
invert this set of linear equations and 
recover the best-fit parameters describing the binned DTD: 
${\bf \Psi}=(\Psi_1,
\Psi_2,...,\Psi_K)$.

In practice, on human timescales SNe in a given
galaxy are rare events ($r_i\ll 1 ~{\rm yr}^{-1}$). Supernova surveys
therefore monitor many galaxies, and record the number of SNe
discovered in every galaxy. For a given model  DTD,
${\bf \Psi}$, the $i$th galaxy
will have an expected number of SNe
\begin{equation}
\label{riti}
\lambda_i=r_i t_i ,
\end{equation}
where $t_i$ is the effective visibility time 
(often called the ``control time'')
during which a SN of a particular type 
would have been visible (given the actual on-target 
monitoring time, the distance to the galaxy, the flux limits of the
survey, and the detection efficiency). Since $\lambda_i \ll 1$,
the number of SNe observed in the $i$th galaxy, $n_i$, obeys a Poisson
probability distribution with expectation value $\lambda_i$, 
\begin{equation}
P(n_i|\lambda_i)=(e^{-\lambda_i}\lambda_i^{n_i})/n_i ! ,
\end{equation}
where $n_i$ is 0 for most of
the galaxies, 1 for some of the galaxies, and more than 1 for very few
galaxies. 

\subsection{Maximum-likelihood DTD recovery}
 We now introduce a nonparametric, maximum-likelihood method
to recover the DTD and its uncertainties. Considering a set of model DTDs, 
 the likelihood of a particular DTD, given the set of measurements
 $n_1,...,n_N$, is
\begin{equation}
L=\prod_{i=1}^N P(n_i|\lambda_i).
\end{equation}
More conveniently, the log of the likelihood is
\begin{equation}
\ln L =\sum_{i=1}^N \ln
P(n_i|\lambda_i)
=-\sum_{i=1}^N{\lambda_i}
+\sum_{i=1}^N\ln (\lambda_i^{n_i}/n_i !), 
\end{equation}
where obviously only galaxies hosting SNe contribute to the second term.
The best-fitting model can be found by scanning the parameter space
of the vector ${\bf \Psi}$ for the value that maximizes the log-likelihood.
This procedure naturally allows restricting the DTD to have
only positive values, as physically required 
(a negative SN rate is meaningless).

The covariance matrix $C_{jk}$ of the uncertainties in the best-fit parameters 
can be found (e.g., Press et al. 1992) by
calculating the curvature matrix,
$$
\alpha_{jk}=\frac{1}{2}\frac{\partial^2\ln L}{\partial\Psi_j \partial\Psi_k}
=\sum_{i=1}^N\frac{\partial[\ln P(n_i|\lambda_i)]}{\partial\Psi_j}
\frac{\partial[\ln P(n_i|\lambda_i)]}{\partial\Psi_k}
$$
\begin{equation}
=\sum_{i=1}^N t_i^2(n_i/\lambda_i-1)^2 m_{ij} m_{ik} ,
\label{curvature}
\end{equation}
and inverting it,
\begin{equation}
[C]=[\alpha]^{-1} .
\end{equation}
Because the values of the DTD are constrained to be positive, if the
maximum-likelihood value of a DTD component, $\Psi_j$, is close to zero, 
the square root of its variance, ${\sqrt{C_{jj}}}$, will not represent
well its  $1\sigma$ uncertainty range. An alternative, 
more reliable, procedure
is to perform a Monte-Carlo simulation in which many mock surveys 
are produced, each having the same galaxies, 
SFHs, and visibility times as the real survey,
and having expectation values $\lambda_i$ based on the best-fit DTD, but
with the number of SNe in every galaxy, $n_i$, drawn from a Poisson
distribution according to $\lambda_i$. The maximum-likelihood DTD,
${\bf\Psi}$, is found for every realization. From the distribution 
 of the values of every component, $\Psi_j$, over 
all the realizations, one can estimate the range encompassing, say,
$\pm 34\%$ of the cases.

The above approach for recovering the DTD has several advantages over
previous methods. First, all of the known information in the survey is
included in the analysis in a statistically rigourous way, including
the fact that many (usually most) of the galaxies did not host any
SNe. Furthermore, 
the calculation is easily generalized to cases where the galaxies
are not all at the same distances (e.g., combinations of surveys
done at different redshifts) -- one simply needs to use the appropriate 
SFH bins and visibility times for every galaxy. 
In fact, assuming that the DTD is a universal function, 
it is straightforward
 to include in a single analysis the data from completely
disparate SN surveys. For example, one could combine the
results of normal SN surveys with
unconventional SN ``surveys,'' in which
the SN rate is measured based on SN remnants
 in small subunits of a few nearby galaxies (Maoz \& Badenes 2010).   

The number and resolution of the time bins used in the analysis will
naturally depend on the quality of the data. 
Larger numbers of observed SNe, $N_{\rm tot}$,
as well as better data on the parent stellar populations (integrated 
colors and/or spectra), 
 will permit a larger number of independent 
SFH and DTD time bins, and will thus improve the time resolution
of the recovered DTD.  We quantify this in \S 4, below.

\section{Application to the LOSS-SDSS Sample}

We now apply our method to the SN survey data obtained by 
considering all LOSS galaxies with SFHs
 based on spectroscopy from the Sloan Digital Sky Survey (SDSS). 
First, we summarize
briefly the essentials of each of these surveys and of the sample
resulting from their intersection.

\subsection{The Lick Observatory SN Search}
 
The LOSS is an ongoing survey for SNe in a sample of $\sim$15,000
nearby (redshift $z < 0.05$) galaxies, conducted with the Katzman
Automatic Imaging Telescope (KAIT) at Lick Observatory (Li et al. 2000;
Filippenko et al. 2001; Filippenko et al. 2010). KAIT is a fully
robotic telescope whose control system checks the weather and performs
observations with a dedicated CCD camera without human intervention.
The data are automatically processed through an image-subtraction
pipeline, and candidate SNe are flagged and visually inspected.
The promising SN candidates are reobserved and the confirmed SNe are
reported to the Central Bureau of Astronomical Telegrams.

A series of papers (Filippenko et al. 2010; Leaman et al.
2010; Li et al. 2010a,b) present the details and first results on
the rates of SNe in the local Universe based on LOSS,
using a sample of 1036 SNe discovered in more than 2 million 
observations between March 1998 through the end of 2008. This
is the largest and most homogeneous set of SN statistics
ever assembled for the determination of local SN rates.
Filippenko et al. (2010) describe the instrumentation and the details
of the SN survey. Leaman et al. (2010, Paper I in this series) present the
control-time calculation for the galaxies in the sample,
for SNe of different types and their luminosity functions (LFs),
and details of the galaxy and SN samples used in the
rate calculations. Monte-Carlo simulations are used to determine
the limiting magnitude and the SN detection efficiency in each LOSS
search image. 

Li et al. (2010a, Paper II in the series) discuss the observed SN LF
using a volume-limited sample ($D < 60 $~Mpc for CC~SNe and $D < 80$~Mpc for
SNe~Ia) of 177 SNe, each with detailed spectroscopic
classification and peak magnitude from dedicated photometric follow-up images
or unfiltered survey images. These observed LFs solve two issues that
have plagued historical SN-rate calculations -- the intrinsic luminosity
distribution of SNe and the host-galaxy extinction. Finally, Li et al.
(2010b, Paper III) combine all of the above ingredients, obtaining
control times for different types of SNe and for each
galaxy, based on its monitoring history, the observed LFs, and the
limiting magnitudes and detection efficiencies of the search images.
These are used to derive SN rates for SNe of different types,
as a function of various galaxy properties.

\subsection{VESPA star-formation histories of SDSS galaxies}
The SDSS (York et al. 2000) 
is a survey of $\sim 10^4~{\rm deg}^2$ of the north
Galactic cap, consisting of imaging 
in five photometric bands ($u, g, r, i, z$), and $3''$-aperture  fibre
spectroscopy of $\sim 10^6$ targets, mostly galaxies, with $r\lesssim 18$~mag. 
Tojeiro et al. (2009) performed spectral synthesis modeling of 
all galaxy spectra in the SDSS using their VESPA code (Tojeiro et al. 2007).  
VESPA uses all of the available absorption
features, as well as the shape of the continuum,
to deconvolve the observed spectra and obtain an 
estimate of the SFH. 
In order to recover the maximum amount of reliable information,
the number of time bins used 
is variable, and depends on the quality of the data
on each galaxy. At the highest resolution, VESPA uses
16 age bins, logarithmically spaced between
0.002 Gyr and $t_0=13.7$~Gyr, the age of the Universe. 
When data do not have sufficiently high signal-to-noise ratio
for a fully resolved reconstruction, 
pairs of adjacent time bins are averaged.
This process may be repeated down to the last two remaining bins.
In the end, the SFH of every galaxy is computed using a different 
set of time bins.

VESPA masses are calculated 
assuming a Kroupa (2007) IMF; all of our results
will therefore include this assumption implicitly.
Bell et al. (2003) have shown that the Kroupa IMF gives a
similar total stellar mass to that of a ``diet Salpeter''  IMF,
obtained by multiplying by 0.7 the total mass of the original 
Salpeter (1955) IMF, to account for the reduced number of low-mass
stars. Thus, our results will be comparable to other SN rate studies,
such as Mannucci et al. (2005, 2006),
that have assumed the diet-Salpeter IMF.

\subsection{The LOSS-SDSS-VESPA sample}

With the kind assistance of R. Tojeiro, we have derived
VESPA SFHs for all of the LOSS galaxies that have SDSS spectroscopy.
Our main sample of LOSS galaxies with VESPA SFHs consists of 3505
galaxies that hosted 201 SNe, among them 82 SNe~Ia, 93 SNe~II,
and 26 SNe~Ibc (see Filippenko 1997 for a review of SN types; here 
we classify SNe~Ib and Ic as ``Ibc'').
An alternative sample, using a different VESPA dust model (see below), 
has slightly different numbers: 3508
galaxies hosting 202 SNe, among them 81 SNe~Ia, 94 SNe~II,
and 27 SNe~Ibc.
These numbers of SNe are consistent with those expected based on 
the fraction of the total LOSS visibility time that is in the VESPA galaxies.
Other than
the increase in Poisson errors due to the smaller number of SNe, the
limitation of the calculation to VESPA galaxies 
should have no effect on our DTD reconstruction.

For the majority of these galaxies, it is practical to separate
the SFHs into no more than four time bins:
$0-70$~Myr, $70-420$~Myr, 
420~Myr$-$2.4~Gyr, and $>2.4$~Gyr. These correspond to the bins labeled 
24, 25, 26,
and 27 in Tojeiro et al. (2009). However, for each galaxy,
the distribution of mass
between the first two bins depends strongly on the particular dust
model
that is assumed, as well as on the spectral synthesis code that is
used -- 
Bruzual \& Charlot (2003), or Maraston (2005) 
(R. Tojeiro, 2009, private communication). In fact, for some models, only a
small fraction of the galaxies have any
 mass formed in the second time bin, a problem that persists at some level 
in other models as well. 
Due to this degeneracy, we have chosen to combine
the first two time bins into a single bin, of $0-420$~Myr.
As the choice of stellar population model parameters affects the
results in other time bins as well, we have used two alternative
VESPA SFH reconstructions. One assumes a single dust component for
each galaxy, while the other allows separate dust components for the
young and the old stellar populations, effectively introducing an
additional free parameter to the modeling (see Tojeiro et al. 2009,
for details).
Both reconstructions use the Maraston (2005) spectral synthesis models.
Using these different SFH reconstructions gives some
idea of the systematic
errors in the DTD reconstruction  arising from the uncertainty in the SFH
(until now, we have treated the SFH values, $m_{ij}$, as error-free 
independent variables).

\subsection{The core-collapse-SN DTD}

\begin{figure}
\includegraphics[width=0.5\textwidth]{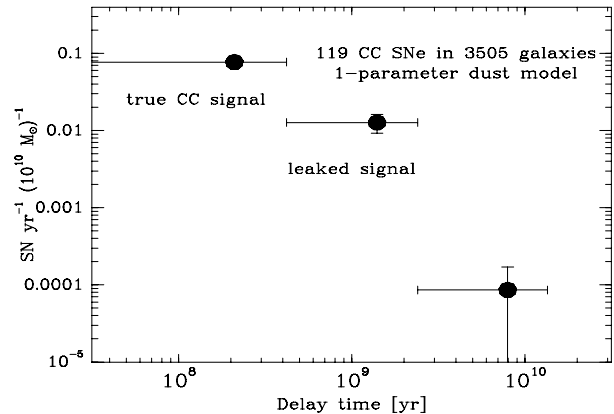}
\caption{Best-fit delay-time distribution
 found for the full LOSS-SDSS core-collapse SN sample using the VESPA SFH 
reconstruction with a single dust parameter. 
Points mark the
best-fit values for each time bin, whose range is indicated by the horizontal
 error bars. Vertical error bars show the most likely 68\% range.
 Note the nonzero amplitude of the DTD in 
the delayed bins, due to ``leakage'' from the prompt bin resulting from 
incorrect characterization of the full stellar populations of the
 galaxies
by the limited aperture of the 
SDSS spectra. 
}
\label{bestfitcc}
\end{figure}
\begin{figure}
\includegraphics[width=0.5\textwidth]{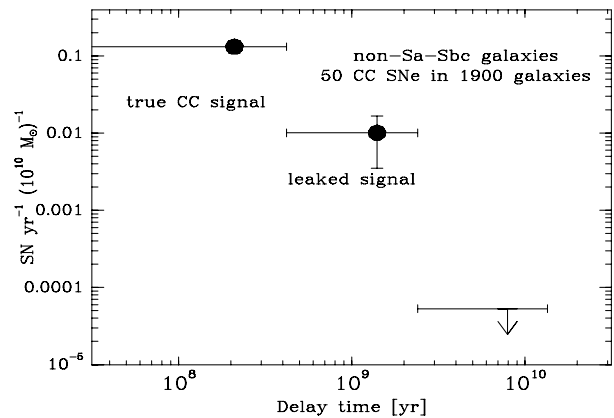}
\caption{
Same as Fig.~\ref{bestfitcc}, but using 
a subsample obtained by culling the LOSS-SDSS sample
of all galaxies of Hubble type Sa--Sbc, for which SDSS fibre spectra do
 not represent well the the full stellar population of a galaxy. 
Symbols are as
in Fig.~\ref{bestfitcc}. Note the enhanced ratio of rates between 
the first and second bins, due to reduced leakage. The best-fit rate
in the third bin is zero, and its 95\% confidence upper limit is marked.
Nevertheless, the leakage is not eliminated completely, and the
statistical errors are larger because of the reduced sample.
}
\label{bestfitccculled}
\end{figure}

We begin by deriving the DTD of the 
CC~SNe in the LOSS-SDSS sample. Core-collapse SNe explode
within $\lesssim 40$~Myr of star formation, and therefore their DTD should
have zero amplitude on timescales much longer than this. This
can provide a first test of our sample and of the DTD recovery method.  

\subsubsection{Full-sample core-collapse SN DTD} 
Figure~\ref{bestfitcc} shows the reconstructed CC~SN DTD, and its
uncertainties from the Monte-Carlo 
simulations described above, for the full sample of 3505 
LOSS galaxies with VESPA SFHs using a
single dust parameter. The SNe in this sample consist
of 93 SNe~II and 26 SNe~Ib and Ic. Since SNe~II and SNe~Ibc 
have different visibility times, we have derived the DTD for
each subsample separately, added the resulting DTD amplitudes
in the corresponding bins, and added the uncertainties in quadrature.
Like the SFHs, the 
DTD in Figure~\ref{bestfitcc}  has three time bins, 
corresponding to $<420$~Myr
(``prompt'' SNe), 420~Myr $-$ 2.4~Gyr (``medium delay'' SNe), and
$>2.4$~Gyr (``delayed'' SNe). The best-fit values and the
uncertainties for these DTDs,
as well as additional ones discussed below, are listed in
Table~\ref{table1}. 


\begin{table*}
\begin{minipage}{150mm}
\caption{Summary of DTD reconstructions}
\begin{tabular}{l|c|c|c|c|c|c}
\hline
\hline
{Sample} &
{$N_{\rm gal}$} &
{$N_{\rm SN}$} &
{$\Psi_{1}$} &
{$\Psi_{2}$} &
{$\Psi_{3}$} &
{$N_{\rm SN}/M$} \\
{} &
{} &
{} &
{0--0.42~Gyr} &
{0.42--2.4~Gyr} &
{2.4--14~Gyr} &
{} \\
{(1)} &
{(2)} &
{(3)} &
{(4)} &
{(5)} &
{(6)} &
{(7)} \\
\hline
CC~SNe&&&&&&\\
\hline
Full CC &3505&119&$770\pm100$&$127\pm35$&$8.6\pm8.5$&$0.00585\pm0.00085$\\
No Sa--Sbc&1900&50&$1320\pm230$&$101\pm66$&$<5.3$&$0.00750\pm 0.00150$\\
$D<100$~Mpc&1951&92&$1037\pm 146$&$192\pm62$&$<4.3$&$0.00920\pm0.00130$\\
$D<65$~Mpc&851&45&$1430\pm340$&$223\pm165$&$<13.4$&$0.01040\pm0.00360$\\
\hline

SNe Ia&&&&&&\\
\hline
Full, 1-dust&3505&82&$84^{+54}_{-39}$&$37^{+22}_{-14}$&$2.6^{+0.8}_{-0.6}$&$0.00137\pm0.00032$\\
Full, 2-dust&3508&81&$26^{+16}_{-20}$&$20^{+24}_{-12}$&$3.3^{+0.8}_{-0.5}$&$0.00086\pm0.00024$\\
No Sa-Sbc
&1900&49&$136^{+110}_{-56}$&$55^{+40}_{-28}$&$3.0^{+1.5}_{-0.6}$&$0.00200\pm0.00060$\\
\hline
\end{tabular}

{Column header explanations: \\
(1)- Sample used to derive DTD: 
{\it Full CC} -- full core-collapse (types II and Ibc) SN sample
with
VESPA SFH reconstructions using a single-parameter dust model;
{\it No Sa--Sbc} -- sample excluding all galaxies of Hubble type
Sa through Sbc, to avoid small-fibre aperture effects;
$D<100$~Mpc -- core-collapse SN sample in galaxies within  $100$~Mpc;
$D<65$~Mpc -- core-collapse SN sample in galaxies within  $65$~Mpc.
{\it Full, 1-dust} -- full SN~Ia sample, with
VESPA SFH reconstructions using a single-parameter dust model;
{\it Full, 2-dust} -- full SN~Ia sample, with
VESPA SFH reconstructions using a two-parameter dust model.
\\
(2) - Number of galaxies in sample.\\
(3) - Number of SNe in sample.\\
(4-6) - DTD rates and 68\% uncertainty ranges, 
in units of $10^{-4}$~SNe~yr$^{-1}(10^{10}\,{\rm M}_\odot)^{-1}$. 
Upper limits, where
given, correspond to a best fit of 0, and to the 95\% confidence
limit.\\
(7) - Time-integrated DTD, in units of SNe~${\rm M}_\odot^{-1}$.
}
\label{table1}
\end{minipage}
\end{table*}

The prompt, $<420$~Myr bin,
in which all CC~SNe should reside, indeed 
has a clear, $>7\sigma$ signal\footnote{We will henceforth use the
  phrasing ``$X\sigma$ signal'' to denote the number of $-1\sigma$ 
($-34$\%) errors by which a best-fit DTD level is above zero. Naturally,
in some cases this will not correspond to the Gaussian probability 
associated with $X\sigma$, either for obtaining zero given the
best-fit result, or for obtaining the best-fit result given zero. We
will discuss these subtleties where relevant.}.
However, there is also a strong $3.5\sigma$-level DTD signal in the 
$420$~Myr$-$2.4~Gyr bin, where no CC explosions are expected.
As we  will demonstrate and quantify 
in \S\ref{crosstalk} below, this result is
 due to the following systematic error.  
The SDSS
spectra of many of the  LOSS galaxies (which are 
relatively nearby, and hence large in angle) are
dominated by the old populations in the centre of each galaxy, due to the 
limited $3''$ aperture size of the SDSS fibres. However,
many of the CC~SNe explode in the outer regions, where there is ongoing star
formation that is invisible to the SDSS spectroscopy. As a result, our
DTD solution mistakenly associates these CC~SNe with an old population,
and hence a large delay. 
We also note that the integrals over these two bins of the best-fit DTD
are comparable: $\Psi_1 \Delta t_1=0.00325$ and $\Psi_2 \Delta t_2
=0.00251$ CC~SNe per M$_\odot$ formed, respectively, where $\Delta
t_1=420$~Myr and $\Delta
t_2=1.98$~Gyr are the lengths of the time bins.
Thus, about 40\% of the prompt-bin signal has leaked into bin 2.
It is unavoidable that the same process will distort the results for
the DTD of SNe~Ia in the same way. 

In the third bin, $>2.4$~Gyr, the best-fit 
amplitude is again nonzero, 
but only at the $1\sigma$ level, and the 
integral over this bin, 
$\Psi_3 \Delta t_3=9\times 10^{-5}\,{\rm M}_\odot^{-1}$, is just a few percent 
of the first two bins. Thus, there is little leakage of power to bin 3.

\subsubsection{Subsamples selected to reduce crosstalk}  
\label{crosstalk}
To understand the effect of the small-aperture SDSS fibres, and
in an attempt to find a method of obtaining cleaner signals in the DTD, we have
experimented with culling the galaxy sample in various ways, based on
the galaxy properties. For example, we expect that much of the ``leak''
that we see is caused by early-type spirals, in which the spectrum
of the bulge, covered by the SDSS fibre, is highly unrepresentative of
the galaxy disk. In contrast, elliptical galaxies and late-type
spirals both have more spatially uniform stellar populations. 
We have therefore derived 
the DTD, shown in Figure~\ref{bestfitccculled}, for CC~SNe from a subsample 
that excludes all LOSS galaxies with Hubble
types 3 to 5 in the notation of Leaman et al. (2010), corresponding
approximately  to  types Sa to Sbc. This reduces the sample size to 1900
galaxies, hosting 
34 SNe II and 16 SNe Ibc (see Table~\ref{table1}). 
As expected, the prompt
signal in $\Psi_1$ is strengthened by a factor 2, while 
$\Psi_2$ is reduced in amplitude. The best-fit value
for $\Psi_3$ is reduced to 0. 

Unfortunately, as seen in Figure~\ref{bestfitccculled} and
in the numbers in Table~\ref{table1}, 
hand-in-hand with the 
increase/decrease in DTD amplitude of the first/second bin,
the smaller SN numbers in the
reduced sample lead to
larger statistical errors. 
We have experimented with additional sample culling methods.
For example, the SDSS database gives, 
for each galaxy, the magnitude
 in each photometric band in an aperture of the size
of the spectrograph fibre, and various measures of the total magnitude
of the galaxy. One can thus select to include in the DTD analysis
only galaxies with no or weak radial color gradients.
Additional or alternative cuts can be made based on galaxy size,
distance, or fraction of total light within the fibre aperture (i.e.,
degree of concentration). 
We find that, as with the case of the selection (above)
according to Hubble type, the more stringent the selection, the lower
the leak out of the first bin in the CC~SN DTD. At the same time, the
lowered sample size increases the statistical errors.

\subsection{The SN~Ia DTD}
Keeping in mind the systematics evidenced in the case of the CC~SNe,
we now proceed to derive the DTD for the SNe~Ia in the sample.   
\begin{figure}
\includegraphics[width=0.5\textwidth]{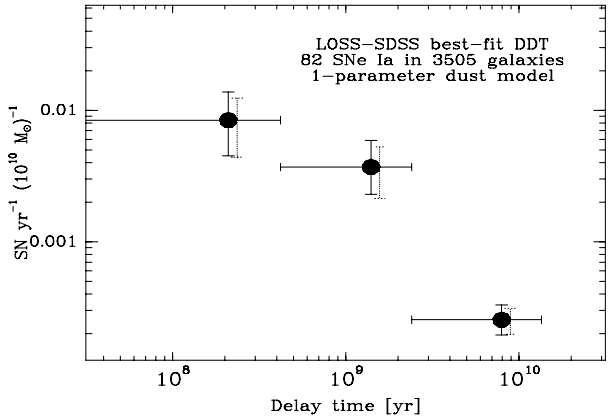}
\caption{Best-fit SN~Ia delay-time distribution
 found for the LOSS-SDSS sample using the VESPA SFH 
reconstruction with a single dust parameter. Points mark the
best-fit values for each time bin, whose range is indicated by the horizontal
 error bars. Solid vertical error bars show the most-likely 68\% range, 
based on Monte-Carlo simulations that use the best-fit DTD values. Dashed 
vertical error bars (slightly shifted to the right, for clarity) show the 
Gaussian $1\sigma$ errors from calculation of the covariance of the 
parameters. The Gaussian errors are symmetric about the best fit, but appear
 asymmetric in the plot because of the logarithmic scale.}
\label{bestfit37}
\end{figure}
\begin{figure}
\includegraphics[width=0.5\textwidth]{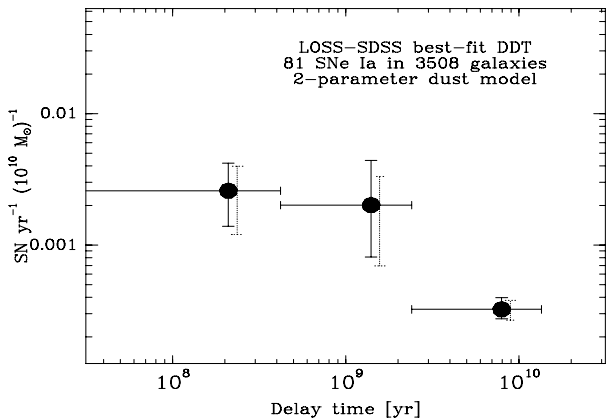}
\caption{Same as Fig.~\ref{bestfit37}, but using a VESPA SFH reconstruction 
permitting separate dust components for the young and  the old stellar populations
of each galaxy.} 
\label{bestfit48}
\end{figure}
Figures~\ref{bestfit37}--\ref{culledbestfitIa} show the 
reconstructed SN~Ia DTDs, and their uncertainties, for the LOSS-SDSS sample,
using the two different VESPA-based SFHs. 
As for the CC~SNe, the best-fit values and the uncertainties for these DTDs
are listed in
Table~\ref{table1}.

\subsubsection{Full-sample SN Ia DTD, one-parameter dust model} 
In the DTD based on the SFHs with a 
one-parameter dust model, we obtain,
in the ``prompt'' ($<420$~Myr) bin, a best-fit value of
$\Psi_1=0.0084^{+0.0054}_{-0.0039}$ SNe yr$^{-1}  (10^{10}\,{\rm M}_\odot)^{-1}$. 
(In other words,
following a $\delta$-function starburst that forms
a stellar mass
of $10^{10}\,{\rm M}_\odot$, the mean SN~Ia rate in this
stellar population
over the first 420~Myr will be $0.0084^{+0.0054}_{-0.0039}$ SNe yr$^{-1}$.)  
This result implies a 
$>2\sigma$ detection of a prompt SN~Ia component. 
As before, the uncertainties we quote are the statistical errors, as derived from the
most probable $\pm 34$\% range in Monte-Carlo simulations, using the 
best-fit values as input. A model with the above best-fit value of $\Psi_1$
as input
yields a recovered $\Psi_1$ of 0 in fewer than 3\% of the Monte-Carlo
realizations. Conversely, a model with an input value of  
$\Psi_1=0$ yields a recovered $\Psi_1 \ge 0.0084$ (in the same units as
before) in $<0.2\%$ of the 
realizations. Models with an input value of  
$\Psi_1\le 0.0031$ yield a recovered $\Psi_1\ge 0.0084$ in $<5\%$ of Monte-Carlo 
realizations. Thus $\Psi_1= 0.0031$ can be considered a
95\%-confidence lower limit on the level of a prompt SN~Ia component.

As another test of the presence of a prompt SN~Ia component, we
have forced $\Psi_1=0$ in the DTD reconstruction. Compared to the
previous result, the log of the likelihood of the best-fit model decreases
by 3.0.
Since $2\Delta \ln L\approx \Delta \chi^2$,
this indicates that a prompt component in the model improves the fit 
to the data at the $2.5\sigma$ level. 
Thus, this test supports, at the $\sim 99\%$
confidence
level, the existence of SNe~Ia
that explode within $420$~Myr after star formation.
 
Since the uncertainties here are larger than 
was the case with the CC~SN DTD, we can use this case to compare
the adequacy of the errors calculated directly using
the covariance-matrix formalism to those from the Monte-Carlo simulations.
 From Figs.~\ref{bestfit37}--\ref{bestfit48}, we see that the analytic
 $1\sigma$ errors
generally match well the errors estimated from the simulations.
 From these simulations, we also find that the
likelihood of the best-fit model is attained in at least 10\%
of simulated trials, indicating that the best-fit model is acceptable  
in an absolute sense as well.

The mean rate found above in the first bin, $\Psi_1$, can be
translated into an equivalent value of the ``$B$'' parameter
(Scannapieco \& Bildsten 2005), the constant of
 proportionality between the SFR   
and the ``prompt'' SN~Ia rate 
(i.e., $B$ is  the number of prompt 
SNe per unit stellar mass formed). Multiplying $\Psi_1$ 
by 420~Myr to integrate 
over the bin, and dividing by $10^{10}$ to express the result per 
solar mass, the best-fit DTD value in the prompt bin implies $B\approx
(3.5\pm
1.7 )\times
10^{-4}\, {\rm M}_\odot^{-1}$ (or $B<8\times
10^{-4}\, {\rm M}_\odot^{-1}$ for the $2\sigma$ upper limit). 
This is several times lower
than the values of $B$ estimated by studies that compare
the SN~Ia rate per
unit mass
and the SFR in blue, vigorously 
star-forming galaxies [see recent
summary and intercomparison in Maoz (2008); observables that are quoted 
there per unit
stellar mass formed need to be divided by 0.7 to convert from the pure
Salpeter IMF assumption to the low-mass-truncated or ``diet'' IMFs
considered here -- Kroupa (2007) or, equivalently in terms of integrated mass,
diet-Salpeter (Bell et al. 2003); see \S3.2]. Such studies have usually 
found values of  $B=(1-3)\times
10^{-3}\, {\rm M}_\odot^{-1}$ (an exception is the value $B=(3.9\pm 0.7)\times
10^{-4}\, {\rm M}_\odot^{-1}$ found by Sullivan et a. 2006). 

As we  showed
in \S\ref{crosstalk} above, at least part of
this discrepancy must be due to the  
limited $3''$ aperture size of the SDSS fibres. 
Just as our CC~SN 
DTD solution mistakenly associated CC~SNe with an old population,
and hence a large delay, it will do so for prompt SNe~Ia that are
associated with star-forming regions in the outer parts of a galaxy,
outside the bulge-light-dominated fibre aperture. In fact, the leak
should be comparable to the 40\% effect we saw for CC~SNe. Thus,
the true $\Psi_1$ level in Figure~\ref{bestfit37} is likely about double
the derived, ``leaky'' one, and thus coincident, within the uncertainties,
with the $B$-parameter values found by other studies, above.
By the same token, our $>2\sigma$ significance on the detection of
a prompt component is a lower limit.

Another contribution to the  difference between the time-integrated
$\Psi_1$ value and published $B$ values may be that
the $B$
parameter found from SN rate measurements in galaxies with the highest 
SFRs are tracing a SN~Ia population with an
even smaller delay time, of $\lesssim 100$~Myr (Mannucci et al. 2005,
2006; Aubourg et al. 2008). The fact that our first time bin goes up
to 420~Myr could then lead to a dilution of $\Psi_1$, due to its being 
averaged over the lower rates at these larger delays.

Turning to the later bins in the recovered SN~Ia DTD, the intermediate-delay
bin of 420~Myr to 2.4~Gyr has a signal of   
$\Psi_2=0.0037^{+0.0022}_{-0.0014}$ SNe yr$^{-1}  (10^{10}\, {\rm M}_\odot)^{-1}$.
Although the uncertainty is still relatively large, there is a $\sim
2.5\sigma$ DTD signal above zero in this bin. Naturally, some or all of
this signal could be due to the ``leak'' from the prompt bin,
discussed above. Based on the observed 40\% leak in the CC~SN DTD,
$\sim 20$\% of the $\Psi_2$ signal could be leaked from $\Psi_1$.  

Finally, the last DTD bin, $>2.4$~Gyr, has the clearest signal,
$\Psi_3=(2.55^{+0.75}_{-0.60})\times 10^{-4}$ SNe yr$^{-1}  
(10^{10}\,{\rm M}_\odot)^{-1}$. At face value, this implies a $\sim 4 \sigma$
  detection of a population of SNe~Ia with large delays. 
The ``$A$'' parameter defined by 
Scannapieco \& Bildsten (2005) was meant to measure the
SN~Ia rate per unit mass in an old population that has no ongoing star
formation. Actually, 
SN~Ia rates
are expected to vary significantly among quiescent
stellar populations of different ages, and therefore like the
$B$ parameter, $A$  will depend on the ages of the stellar populations
probed by a SN survey. Indeed,
the typical values found for the $A$ parameter
 have been in the range $A \approx (2-10)\times 10^{-4}$ SNe yr$^{-1}  
(10^{10}\,{\rm M}_\odot)^{-1}$ (see compilation in Maoz 2008). 
However, $\Psi_3$ is the rate per unit mass {\it formed}. The published
rates in old populations are a per unit of {\it existing} stellar mass 
(i.e., in stars and stellar remnants). 
The difference is given by the stellar mass returned to the 
interstellar medium via SN explosions and mass loss during 
stellar evolution. The fraction of returned material is an increasing
function of time, and corresponds to $\sim$25\% after 0.1 Gyr and $\sim$50\%
after 12 Gyr (Bruzual \& Charlot 2003). Thus,
for comparison to the $A$ parameter, $\Psi_3$ needs to be doubled,
giving a rate of
$(5.1^{+1.5}_{-1.2})\times 10^{-4}$ SNe yr$^{-1}  
(10^{10}\,{\rm M}_\odot)^{-1}$, in excellent agreement with other measurements
of SN~Ia rates in old populations.  
In principle,
some or all of the $\Psi_3$ signal could again be the result of a leak from
the prompt bin, due to the limited spatial coverage of the SDDS
fibres. In practice, the time integral over the $\Psi_3$ rate, 
$\Psi_3 \Delta t_3$, is
seen to be comparable to $\Psi_1 \Delta t_1$, as opposed to the 
few-percent leak into the $\Psi_3$ bin that we found in the CC~SN case in 
\S\ref{crosstalk}. This suggests 
that our $\Psi_3$ value is real and largely uncontaminated. 
Furthermore, examining the time integrals
over the best-fit DTD in 
each of the bins, and attempting to correct for the leak from bin
1 to bin 2, suggests a relative contribution to the total SN~Ia
numbers of (prompt:medium:delayed) $\approx$ 2:2:1. However, this is
subject to large statistical and systematic uncertainties.
 
It has been known for a long time that SNe~Ia can occur in early-type
galaxies with little or no star formation, and hence our measurement
of a significant delayed component is hardly revolutionary. However, a
SN~Ia in a particular early-type galaxy can always be attributed to
some residual low-level star formation combined with 
 a short SN~Ia delay time. 
Our measurement, on the other hand, provides a statistically robust 
determination of the delayed component, and its level relative to the
prompt components, in a population with detailed measured SFHs. 

\subsubsection{Full-sample SN Ia DTD, two-parameter dust model} 
Using the second VESPA SFH reconstruction that utilises a two-parameter dust
model, 
the best-fit DTD values (Fig.~\ref{bestfit48} and Table~\ref{table1}) 
for the prompt, medium, and delayed
components are
somewhat different from those obtained
with the one-parameter dust model, However, these systematic 
differences are smaller than  the statistical uncertainties of the
results quoted above. 
For larger SN samples with smaller statistical uncertainties, the
systematic uncertainties due to SFH modeling may, however, become dominant. 
In what follows, we will use only the
VESPA SFHs based on the first, one-parameter dust model.

\subsubsection{Culled-sample SN~Ia DTD}
\begin{figure}
\includegraphics[width=0.5\textwidth]{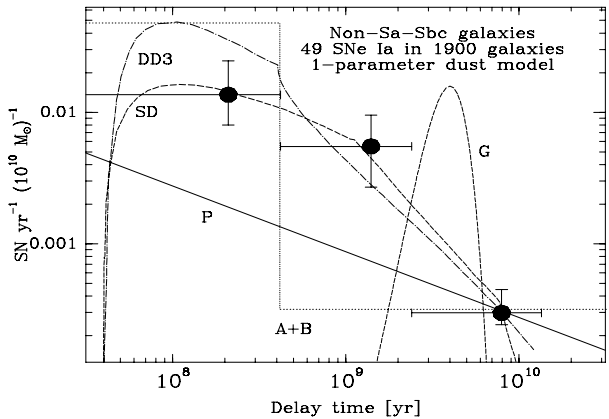}
\caption{Best-fit SN~Ia delay-time distribution
 found for a subsample obtained by culling the LOSS-SDSS sample
of all galaxies of Hubble type Sa--Sbc, for which SDSS fibre spectra do
 not represent well the full stellar population of a galaxy. 
Symbols are as
in Fig.~\ref{bestfit37}. Note the enhanced ratio of rates between 
the first and second bins, due to reduced cross-talk.
Also plotted, for comparison, 
are several empirically or theoretically motivated 
DTD models: {\bf A+B}, representation of 
the Scannapieco \& Bildsten (2005) ``$A+B$'' model, plotted at the median
levels of various estimates, as compiled by Maoz (2008); {\bf G},
 Gaussian DTD centred at 4~Gyr with half-width 0.8~Gyr, proposed by
 Strolger et al. (2004); {\bf P}, $t^{-0.5}$ power law, suggested by
 Pritchet et al. (2008); {\bf DD3}, double-degenerate ``DD-Close-3''
 model by Greggio (2005) with $3M_\odot$ minimum initial mass; and
 {\bf SD}, single-degenerate 
 model of Greggio (2005), as shown by Greggio et al. (2008).
The latter three models have been normalised to go through the
 best-fit 
observed LOSS-SDSS $\Psi_3$ rate for this subsample.
}
\label{culledbestfitIa}
\end{figure}

As in \S\ref{crosstalk},
we have repeated 
the DTD derivation for SNe~Ia from the subsample of 1900 galaxies 
that excludes Hubble
types  Sa to Sbc. This sample hosts 49 SNe Ia. 
Figure~\ref{culledbestfitIa} shows the SN~Ia DTD obtained for this 
culled sample.

As was the case for the CC~SNe in the culled sample, 
the prompt-bin amplitude
is enhanced compared to the full sample DTD, and the medium-delay bin
rate is lowered. The rate in the third bin, in contrast, is hardly changed
compared to Figure~\ref{bestfit37}, reconfirming the
reality of that signal. 

\subsubsection{Comparison to DTD models}
\label{comparisontomodels}

It is beyond the scope of this paper to attempt a 
detailed comparison of our recovered DTD to the many models
that have been proposed, and to address the progenitor issue.
Nonetheless, we have superimposed in Figure~\ref{culledbestfitIa},
which is our most reliable SN~Ia DTD (insofar as the leakage problems
are partly mitigated in it), a selection of DTDs that have appeared in
the literature. Some are empirically motivated, in the sense that they 
were used to relate SN-rate data with star-formation measurements,
using one of the methods outlined in \S1. Others are more
theoretically motivated, based on a progenitor scenario. 

The Scannapieco \& Bildsten (2005) ``$A+B$'' model has been discussed
several times above. It is not exactly a DTD, but rather
a prescription for relating a SN rate to a galaxy of a given current mass
and SFR. Nevertheless, we have overplotted this model 
in Figure~\ref{culledbestfitIa}, using the median literature values
of the $A$ and $B$ parameters compiled by Maoz (2008). We have
associated the prompt $B$ component with the same 420~Myr bin of our
DTD reconstructions, and the level to be plotted is then just 
the median level,  $B=2\times
10^{-3}\, {\rm M}_\odot^{-1}$ divided by this time interval.
The median value for the $A$ parameter is $A =
6\times 10^{-4}$ SNe yr$^{-1} (10^{10}\,{\rm M}_\odot)^{-1}$, 
which we plot as a constant rate at times $>420$~Myr, but halved
to account for stellar mass loss (see above). As already noted,
``$A+B$''  is an oversimplification, but its measured values are seen
to be consistent with our DTD. 

Strolger et al. (2004) and Dahlen et al. (2004) have deduced a 
 Gaussian DTD centred at 4~Gyr with half-width 0.8~Gyr, as a best fit
 to their comparison of SN rates to cosmic SFH, out to $z=1.6$.
As seen in Figure~\ref{culledbestfitIa}, such a DTD is strongly 
at odds with our local, directly derived, DTD. 

 Pritchet et al. (2008) have argued for a power-law DTD form,
 $t^{-0.5\pm 0.2}$, based on analysis of the Supernova Legacy Survey
 data. A dependence of roughly $t^{-0.5}$ is also expected simply
from the formation rate of WDs, when considering the IMF 
and main-sequence stellar lifetimes. 
In Figure~\ref{culledbestfitIa}, we have plotted a $t^{-0.5}$
curve, normalised to pass through the $\Psi_3$ rate, which is
the most robust and stable rate in our DTD reconstruction. 
It appears that a $t^{-0.5}$
dependence is too shallow to match our derived DTD, especially
 considering that $\Psi_1$ is likely underestimated by us, given
the possibility of remaining leakage from $\Psi_1$ to $\Psi_2$, 
even in our culled sample (as evidenced from the CC~SN sample).

Finally, we have overplotted in Figure~\ref{culledbestfitIa} two of the
analytical models of Greggio (2005), as presented by Greggio, Renzini,
\& Daddi(2008),
based on stellar-evolution arguments 
and on various parametrisations of the possible 
results of the complex common-envelope phases through which SN~Ia
 progenitor
systems must pass. 
For each of several SN~Ia channels, Greggio (2005) calculates
 the DTDs
that emerge when varying the values for the parameters describing the
 initial conditions, and the mass and separation distributions and 
limits of the systems that eventually explode.
We show here one SD model and one DD model, again normalised to pass through 
our best-fit $\Psi_3$ rate.

The ``DD-Close-3'' label of the DD model refers
 to one of two possible
 parametric schemes used by Greggio (2005) to describe 
the WD separation distribution after the common-envelope phase, and
to a minimum assumed initial mass of the secondary star in the binary,
 of $3M_{\odot}$. This DD model appears to match well 
our recovered DTD, especially considering that we have residual
leakage problems from $\Psi_1$ to $\Psi_2$, and hence $\Psi_1$ is 
underestimated. It is easy to see from Figure~\ref{culledbestfitIa}
that this model, after its initial rise to maximum, is essentially a 
broken power law. At $t\lesssim 400$~Myr, the slope is $-0.5$, just like the 
Pritchet et al. (2008) model slope. This slope is the result of stellar-evolution 
lifetimes and IMF (see above). At $t>400$~Myr, 
the slope is $-1.3$. A slope of roughly $-1$ is generic to models in which 
the merger rate is determined by energy loss to gravitational waves 
(e.g., Greggio 2005; Totani et al. 2008). 

The Greggio (2005) SD model also matches our 
recovered LOSS-SDSS DTD in Figure~\ref{culledbestfitIa} remarkably
well. It should be remembered, however, that in a full physical model,
the normalisation is not a free parameter, and is dictated by the
efficiency of SN~Ia production from the potential progenitor
population. As emphasised by Maoz (2008), and further discussed 
in \S\ref{sectionIayield}, below, 
the efficiencies found
by binary population synthesis models are at least a factor of a
few, and likely an order of magnitude, lower than indicated by observed
SN rates (see also Mennekens et al. 2010). 
Furthermore, the SD model, whose shape matches our derived 
DTD so nicely, has been found to be even less efficient, by yet
another order of magnitude, in some studies (e.g., Tutukov \& Yungelson 2002).
 
 From this brief comparison of our recovered DTD to previous work, we 
conclude the following. (1) The monotonically decreasing nature of the 
DTD that we have found is 
in agreement with most previous empirical determinations and
theoretical expectations, except for that of Strolger et al. (2004).
(2) The decline with time of our DTD is steeper than in the Pritchet et
al. (2008) model. Two of the Greggio (2005) models, SD and
DD-Close-3, fit well the shape of the recovered LOSS-SDSS DTD.
(3) The normalisation of full physical models to the observations has not 
been considered here, but it is another point at which models are
challenged by these and other observations of SN rates.



\subsection{The time-integrated DTD}

Another interesting observable is  
the integral of the DTD over the age of the universe ($t_0$),
\begin{equation}
N_{\rm SN}/M=\int_0^{t_0}\Psi(t)~ dt ,
\label{intddt}
\end{equation}
which gives the total number of
SNe that eventually explode, per unit stellar mass formed in a short
burst of star formation.
We can approximate the integral in Eq.~\ref{intddt} 
with a sum over the binned DTD,
\begin{equation}
N_{\rm SN}/M\approx\sum_{j=1,K}\Psi_j \Delta t_j.
\label{nmsum}
\end{equation}
The adequacy of this approximation will depend on the true form 
of $\Psi(t)$ and on the number of bins. From the simulations described
in the next section, we find that, for declining power-law DTDs
represented with three time bins, Eq.~\ref{nmsum} typically underestimates
$N_{\rm SN}/M$ systematically by $\sim 10-20\%$. 

\subsubsection{Core-collapse SN yield per stellar mass}
If all
stars with mass $>8\,{\rm M}_\odot$ explode as CC~SNe, then for CC~SNe, 
$N_{\rm SN}/M$ is
 just the ratio of the number of stars formed above this mass limit to the
total stellar mass formed, and is  
easily calculated to be $N_{\rm CC}/M=0.010 ~ {\rm SNe}~
{\rm M}_\odot^{-1}$ for the ``diet'' Salpeter IMF (as well as for 
the Kroupa (2007) IMF assumed by the the VESPA SFH reconstruction). 
In contrast, the integral over our reconstructed
DTD of the CC~SNe 
 is $(0.00585\pm 0.00085)~ {\rm SNe}~
{\rm M}_\odot^{-1}$, a factor of $\sim 2$ lower than expected\footnote{
Here, the statistical 
error on $N_{\rm CC}/M$ is found using the covariance matrix $[C]$
for ${\bf \Psi}$, and the vector of time-bin intervals ${\bf \Delta t}=(\Delta t_1, \Delta
t_2,...,\Delta t_K)$, such that
$  
\Delta(N_{\rm SN}/M)=({\bf \Delta t}[C]{\bf \Delta t})^{1/2}$.}.
However, we can show that this deficit of CC~SNe largely disappears
in various subsamples of the LOSS-SDSS sample, and is therefore
not a real effect. 

\begin{figure}
\includegraphics[width=0.5\textwidth]{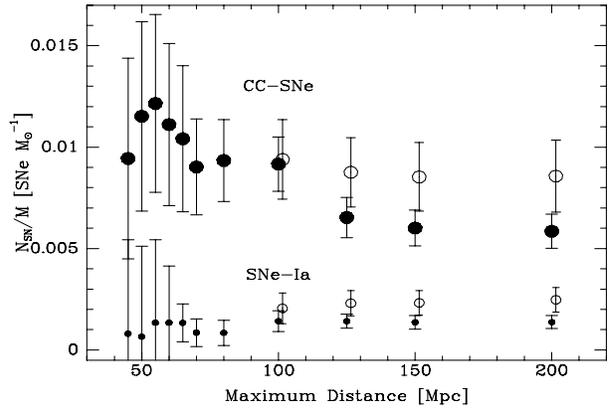}
\caption{The time-integrated DTD, i.e., the total number of SNe
  produced over a Hubble time, per unit stellar mass formed, for 
different distance-limited galaxy subsamples. Top full symbols are for
  CC~SNe (SNe II plus SNe Ibc), bottom full symbols are for SNe~Ia.
Top empty symbols are for CC~SNe in a subsample of galaxies with 
masses $<3\times 10^{10}\, {\rm M}_\odot$.
Bottom empty symbols are for SNe~Ia in a subsample of galaxies with 
masses $<7\times 10^{10}\, {\rm M}_\odot$.
The rise in $N_{\rm CC}/M$ toward 
its theoretically expected value with decreasing maximum sample
distance or with sample culling is another manifestation of the
SDSS small-fibre-aperture problem.
 The value of $N_{\rm CC}/M$ at $\lesssim 65$~Mpc indicates
that most stars above 8 M$_\odot$ produce CC~SNe. The constant value
of $N_{\rm Ia}/M$, on the other hand, shows that the systematics
affecting SNe~Ia are less dependent on distance. Nevertheless,
culling the higher-mass galaxies from the sample increases
$N_{\rm Ia}/M$ at all distances.
The time-integrated ratio of CC~SNe to SNe~Ia is seen to be roughly 4:1.  
} 
\label{nmfig}
\end{figure}

For example, we have reduced the LOSS-SDSS galaxy sample by
limiting the maximum galaxy distance to progressively smaller values.
For each distance-limited sample, we recover the DTD.  
 Figure~\ref{nmfig} shows the values of $N_{\rm CC}/M$ we obtain
by integrating over the reconstructed CC~SN DTDs, for 
each distance limit. Several of these values are also listed in 
Table~\ref{table1}. Clearly, $N_{\rm CC}/M$
 rises with decreasing distance, and
reaches close to the theoretically expected value when the sample is limited to
galaxies closer than $D \approx 65$~Mpc. From this plot we estimate the 
observed limiting value to be
$N_{\rm CC}/M=0.010\pm0.002$. 


Although a potential  explanation for the dependence of $N_{\rm CC}/M$ on
sample volume would be an increasing fraction of missing CC~SNe with distance,
perhaps due to mild obscuration of such SNe by dust, this is unlikely 
to be the correct explanation. $N_{\rm CC}/M$ is derived from the DTD,
which takes
into account the visibility time of each type of CC~SN at the
distance of each galaxy in the sample. The
visibility time, in turn, was calculated, as described in Li et
al. (2010a), using the observed LF of 87 nearby
CC~SNe, in which the effects of extinction are automatically included.

More likely, the decrease of $N_{\rm CC}/M$ with distance is another
manifestation of the SDSS fibre aperture problem. At larger
distances, there are fewer low-mass galaxies 
and late-type spirals in the LOSS sample, and
more high-mass intermediate and early types. 
In the earlier-type galaxies, VESPA
in more likely to overestimate the total galaxy mass, based on the
spectrum of the bulge population probed by the SDSS aperture.
Indeed, if we rederive the CC~SN DTD and $N_{\rm CC}/M$, but exclude 
the earlier Hubble
types (as in \S\ref{crosstalk}), or exclude massive galaxies,
  the values of $N_{\rm CC}/M$ for the
samples limited to $D<125$~Mpc, $D<150$~Mpc, and $D<200$~Mpc rise significantly,
and approach their $D<100$~Mpc values. For example, 
we show in Figure~\ref{nmfig} (upper empty symbols)
that, if we limit the sample 
to galaxies less
massive (based on their VESPA reconstructions)
than $3\times 10^{10}\, {\rm M}_\odot$, the best-fit DTDs
at those distances give $N_{\rm
  CC}/M=0.0086\pm 0.0017$, much closer to the expected value.
(Naturally, the reduced sample sizes lead to larger statistical 
error bars in Fig.~\ref{nmfig}.)
Another indication of the nature of the effect can be seen in Table~1,
where the ratio of $\Psi_2/\Psi_1$, which quantifies the leak from the
first to the second bin, decreases for the CC~SNe as $D$ is reduced.

The fact that a galaxy sample can be defined (the 
$D\lesssim 65$~Mpc sample in the above example), for  which 
$N_{\rm CC}/M$ actually reaches its expected value, is important. 
It confirms that,
indeed, the majority of stars with $>8\,{\rm M}_\odot$ produce CC~SNe. 
This cannot be taken for granted. Given current observational and
theoretical limits, the low-mass limit for core collapse could be as
high as 10 or even $11\,{\rm M}_\odot$ (see Smartt et al. 2009, and references
therein). Moving the limit from 8 to $10\, {\rm M}_\odot$ would
decrease the expected $N_{\rm CC}/M$ by about 30\% (i.e., to our $2\sigma$
observed lower limit on $N_{\rm CC}/M$).
Raskin et al. (2008) have recently obtained a similar estimate of
the lower mass limit for core collapse by matching, on the one hand,
 the differences
in the spatial distributions of stars and SNe in spiral galaxies to,
on the other hand,
the predictions of simple stellar-population aging models. 
  
Furthermore, in principle, a sizable fraction of high-mass stars could
collapse directly to a black hole, with only a weak or null SN
explosion accompanying the collapse (Kochanek et al. 2008).
Our measurements of $N_{\rm CC}/M$ also argue against this scenario,
unless the minimum mass for CC~SNe is even lower than $8\,{\rm M}_\odot$.
We note that these conclusions are reinforced by 
the fact that the discrete sum in Eq.~\ref{nmsum} 
underestimates the true integral over $\Psi$. 

\subsubsection{The SN Ia yield per stellar mass}  
\label{sectionIayield}
Figure~\ref{nmfig} also shows the results of a similar analysis 
for the SNe~Ia: the integral $N_{\rm Ia}/M$ over the 
best-fit SN~Ia DTD
(shown in Fig.~\ref{bestfit37} for the full sample), again as 
a function of the sample distance limit. We see (lower filled symbols) 
that there
is no obvious dependence on distance, with $N_{\rm Ia}/M$ apparently
constant at  $\sim 0.00140\pm 0.00033 ~{\rm SNe}~ {\rm M}_\odot^{-1}$. 
However, if we again limit the sample in mass, for example to 
$<7\times 10^{10} {\rm M}_\odot$, we obtain (lower empty symbols in 
Fig.~\ref{nmfig}) higher values, 
$N_{\rm Ia}/M\approx 0.00230\pm 0.00060 ~ {\rm M}_\odot^{-1}$.
The systematics of the masses of the SN~Ia host galaxies, while
present, are
apparently less dependent on distance than those of the CC~SN hosts.

These values of $N_{\rm Ia}/M$ are also of interest.
 As already noted above, previous studies have found, just for the
 prompt SN~Ia component embodied in the $B$ parameter, 
 values of $B=(1-3)\times 10^{-3} {\rm M}_\odot^{-1}$. We note that 
 the $B$ parameter, like $N_{\rm Ia}/M$,
  relates to the stellar mass {\it formed}.
Mannucci et al. (2006) obtained an empirical DTD based on the
observations described by Mannucci et al. (2005) and 
Della Valle et al. (2005).
Their time-integrated SN rate is $0.0013~{\rm SNe}~ {\rm M}_\odot^{-1}$.
However, this rate is per unit of {\it existing} stellar mass, 
after assuming 
a fraction of recycled gas of 0.30 as an average among populations of 
different ages (Bruzual \& Charlot 2003).
Accounting for this factor, the Mannucci et al. (2006) result
corresponds to $\sim 0.0010~{\rm SNe}~ {\rm M}_\odot^{-1}$
 of {\em formed} stars, similar to the
values measured here.

As discussed at length by Maoz (2008),  $N_{\rm Ia}/M$ is one of
several observables that can be related 
directly to the fraction $\eta$ of stars in some initial mass range $[m_1,m_2]$
that eventually explode as SNe~Ia:
\begin{equation}
\label{frac1}
\eta=\frac{N_{\rm Ia}}{M}\frac{\int_{0.1}^{100} m (dN/dm) dm} 
{\int_{m_1}^{m_2}(dN/dm) dm},
\end{equation} 
where $dN/dm$ is the IMF. For the ``diet Salpeter'' IMF (which, again,
gives results similar to the Kroupa IMF assumed by the VESPA SFH 
reconstruction), and an initial  mass range of 3--8 M$_\odot$, often considered
for the primary stars of SN~Ia progenitor systems, the ratio  of the two
integrals equals 33. Adopting the higher value of $N_{\rm Ia}/M$ that we
have found in the culled sample, under the assumption that it is more robust against the
fibre-aperture effect, we obtain $\eta=7.6\pm 2.0 \%$. 
This is in agreement with 
the results of Mannucci et al. (2006), who found $\eta=4.3\%$ for
these parameters.
As noted by
Maoz (2008), the consistently high values of the exploding fraction, $\eta$, derived from
several different observables, may constitute a problem for
current progenitor models. 

Interestingly,  
in galaxy clusters, Maoz et al. (2010) estimate 
a time-integrated number of SNe~Ia per present-day stellar mass
of $0.011 ~{\rm SNe}~
{\rm M}_\odot^{-1}$. This estimate is based on the
observed ratio of total mass of iron (both in stars and in the
intracluster medium) and the mass in stars. For an IMF with a
 standard high-mass end, 
the present-day mass observed in stars is related to the number of 
CC~SNe, whose contribution to the iron mass has been estimated and
subtracted. The
remaining iron mass is then due to the SNe~Ia. 
Multiplying by 0.5 to convert to formed mass, rather
than present mass, this gives
$N_{\rm Ia}/M=0.006 ~{\rm SNe}~
{\rm M}_\odot^{-1}$. Maoz et al. (2010) show that, given the uncertainties
in observed cluster properties, this number could decrease by perhaps
a factor of 0.6 at most, to
$N_{\rm Ia}/M=0.0035~{\rm SNe}~{\rm M}_\odot^{-1}$. 
This is still at least a factor of 1.5
greater than, but marginally consistent with,
the $N_{\rm Ia}/M=0.0023\pm 0.0006 ~{\rm
  SNe}~{\rm M}_\odot^{-1}$ that we have found here based on LOSS.

The cluster-based value of $N_{\rm Ia}/M$ 
 may be evidence
for early enrichment of clusters by CC~SNe from a top-heavy IMF. 
These CC~SNe from massive stars, which left no traces in the form
of low-mass relatives, would have then produced the bulk of the iron
mass in clusters. Alternatively, the large iron mass in clusters
could have indeed come from SNe~Ia, but this would imply a more
efficient production of SNe~Ia in cluster environments. Intriguingly,
Sharon et al. (2007), Mannucci et al. (2008), and Graham et al. (2008)
have all found evidence for SN~Ia rates 
enhanced by factors of a few in cluster galaxies, compared to 
field early-type galaxies.

\subsubsection{The ratio of CC~SNe to SNe~Ia}  
 From the ratio of the time-integrated DTDs, $N_{\rm CC}/M$ and $N_{\rm
 Ia}/M$ (using the value of $N_{\rm CC}/M$ at $D\lesssim 70$~Mpc
for which the CC~SN counts are fairly complete, but for which the errors
are not excessively large because of the limited sample size, 
and $N_{\rm Ia}/M$ from the low-mass sample, see above), 
the time-integrated ratio of CC~SNe to SNe~Ia is  
$N_{\rm CC}/N_{\rm Ia}=4^{+3}_{-1.5}$. 
If we force the value of ${N_{\rm CC}}/M=0.01$ expected from 
a diet-Salpeter IMF with a low-mass CC limit of 8~M$_\odot$,
the allowed range in the ratio shrinks to
$N_{\rm CC}/N_{\rm Ia}=4^{+2}_{-0.7}$).
We note that the ratio of time-integrated DTDs
 is distinct from the observed ratio
of {\it current rates}, which is measured to be about 
3:1 in local surveys (Mannucci et al. 2005; Li et al. 2010a).
The ratio of current rates depends on the summed SFH of the galaxies in the
 volume. It can be arbitrarily high for very young, star-forming
 populations (in which few SNe Ia have had time to form yet) to zero for
old, inactive populations (with no CC~SNe). In contrast, the ratio of the
 time-integrated DTDs, like the DTDs themselves, is independent of SFH,
and intrinsic to the stellar-evolution processes that lead to SNe.

The ratio $N_{\rm CC}/N_{\rm Ia}$ of course equals $(0.01\, {\rm M}_\odot^{-1})/
(N_{\rm Ia}/M)$, for the diet Salpeter IMF and the said
lower limit for CC~SNe, and thus
the cluster-based lower limit of $N_{\rm Ia}/M>0.0035~{\rm
  SNe}~{\rm M}_\odot^{-1}$ (Maoz et al. 2010), discussed above, 
implies $N_{\rm CC}/N_{\rm Ia}<2.9$,
and possibly even 1:1.
De Plaa et al. (2007),
comparing cluster element abundances to theoretical SN element 
yields, have deduced an 
integrated CC~SN to SN~Ia ratio of 1:1. On the other hand, the
observational uncertainties are such that a time-integrated CC~SN to
SN~Ia 
ratio of roughly
3:1, despite some tension,  
is consistent both with cluster measurements and with the LOSS data we have
analysed here.

\section{Tests on Simulated Samples}

To test our DTD recovery procedure and examine its performance
on different kinds of input datasets, we have repeated the 
Monte-Carlo mock survey generation described in \S2, above, using the
LOSS-SDSS galaxy sample and its VESPA-based SFHs (with one dust
parameter). However,
rather than inputing the best-fit DTD from the inversion of the real data, we
can choose other model DTDs and
convolve those DTDs with $m_{ij}$ according to
Eq.~\ref{discreteconvolution}--\ref{riti}, or we can change the sample
properties in additional ways, as described below. 
Using the actual LOSS galaxy SFHs and visibility times (as opposed to,
for example, random SFHs and visibility times)
makes for a more realistic simulation.
As before,
the number of SNe found in each galaxy in each mock survey is drawn
from a Poisson distribution with expectation value $\lambda_i$. 
The  different-shaped DTDs which can be input and recovered can also be scaled
up or down to produce a larger or smaller total number of SNe in the 
mock survey (or, equivalently, the visibility times, $t_i$, can be scaled up or down).
In these simulations, the SFHs are assumed to be error-free
independent variables, and are used both for creating the mock samples
and deriving their DTD. Therefore, these simulations will not display 
the ``leakage'' between bins that we have encountered with the
LOSS-SDSS sample, nor other problems due to systematic or random 
errors in the SFHs of the galaxies.

 From our simulations, we find that the distribution of
output DTD amplitudes is centered on the input DTD
values, meaning that the method reliably recovers the input DTD with
little bias. Nonetheless, for low DTD amplitudes combined 
with large uncertainties (due to small SN numbers, see below), there 
can be some ``pile-up'' in the distribution
at zero amplitude, as a result of the positivity constraint of the
DTD.
Thus, obtaining a zero amplitude for a bin in the DTD
reconstruction can happen, even when in reality the amplitude is
nonzero but low, and the number of SNe in the survey is small.

We find that
the relative uncertainty
in the DTD amplitude in the $j$th bin scales roughly as
one would expect from Poisson statistics, considering the number of
SNe that contribute to every bin in the DTD. For example, with three
time bins, the relative error in the first bin of the DTD is
\begin{equation}
\frac{\Delta\Psi_1}{\Psi_1}\approx \left(N_{\rm tot}\frac{\sum_{i} m_{i,1}\Psi_1}
{\sum_{i} m_{i,1}\Psi_1+\sum_{i} m_{i,2}\Psi_2+\sum_{i} m_{i,3}\Psi_3}
\right)^{-1/2}.
\end{equation}
For a survey with a fixed total number of SNe,
$N_{\rm tot}$, there will thus be a tradeoff in the analysis between DTD
accuracy and resolution. The uncertainties do not depend on the number of
galaxies monitored, and thus a brief-duration survey of many galaxies
and a long-duration survey of few galaxies are equivalent, as long as
they produce the same total number of SNe.  

\begin{figure}
\includegraphics[width=0.5\textwidth]{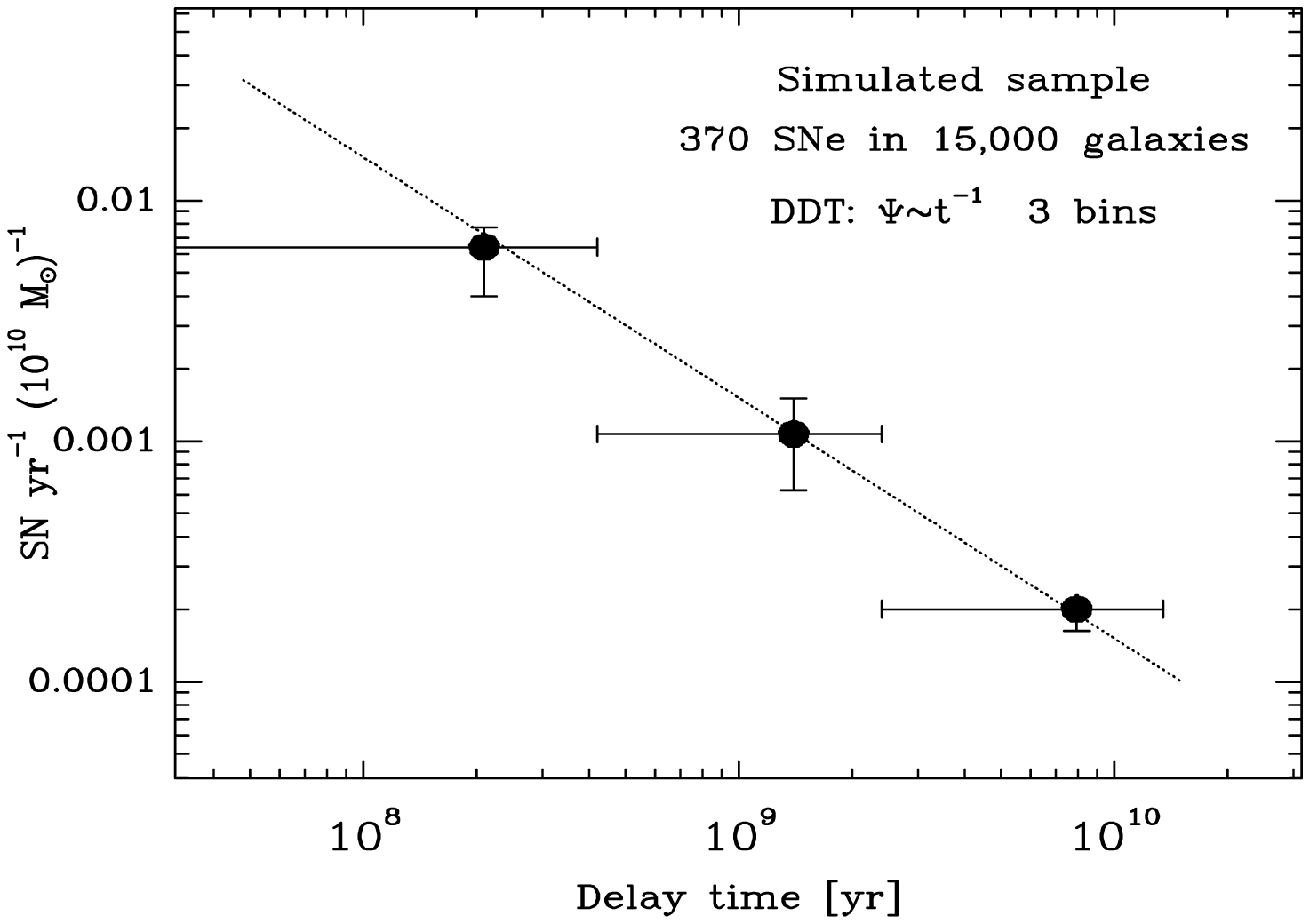}
\end{figure}
\begin{figure}
\includegraphics[width=0.5\textwidth]{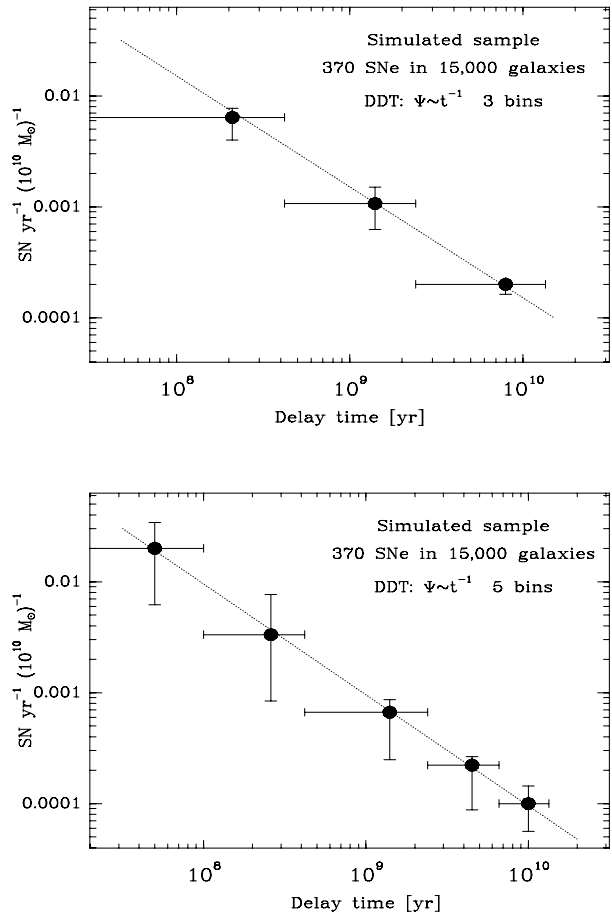}
\caption{DTD recovered using three or five bins 
from a mock sample of 15,000 galaxies hosting 
370 SNe~Ia, the numbers characterising the full LOSS sample. The
input DTD is a $t^{-1}$ power law (dashed line). 
Points are plotted at the value of the input binned DTD, 
and error bars show the 
ranges that include 68\% of the recovered best-fit solutions
in repeated random realizations of SN surveys on the mock sample.}
\label{totalloss}
\end{figure}

Figure~\ref{totalloss} examines the quality
of the reconstruction as the number of time bins is varied between
three and five.
Here, we have assumed a survey with 15,000 galaxies and $\sim 370$
SNe, similar to the full LOSS SN~Ia sample, if all galaxies in it had SFH 
reconstructions available. To produce this large mock sample, we simply
clone several times the 3505 SFHs and visibility times of the LOSS-SDSS
sample, with its three time bins. To obtain a sample with five time bins 
in each galaxy's SFH, we have split, in this example, the first SFH
bin into two bins: a 0--100~Myr bin and a 100--420~Myr bin. The stellar
mass in each sub-bin was randomised by $\pm 50$\% around the 
mass value corresponding to the relative time fractions. (Such
randomisation is essential, as otherwise the $m_{ij}$ are no longer
independent variables.). Similarly, we further
split the last bin into two bins, corresponding to 2.4--6.5~Gyr and
6.5--13.5~Gyr.
For the DTD in these simulations, we have taken a
$\Psi(t)\propto t^{-1}$ dependence (which is roughly generic to DD
models; see \S\ref{comparisontomodels}, above), scaled so as to give the
desired total number of SNe using the LOSS visibility times. 
The simulations show that
surveys with several hundred SNe enable 
reasonably good DTD recovery in terms of both accuracy and temporal resolution.
Specifically, Figure~\ref{totalloss} demonstrates that already-existing SN
surveys, and LOSS in particular, have the power to measure reliably
the SN~Ia DTD,
{\it if} the SFHs could be estimated in a comprehensive and unbiased way
for the full sample of 
$\sim 15,000$ galaxies. We briefly discuss the prospects for 
this in the concluding
section, below.

We have gauged the effect of the time binning on the time integral 
over the DTD, $N_{\rm SN}/M$, 
as approximated in Eq.~\ref{nmsum}. To do this, we
create mock samples using large numbers of bins in the SFH and the
DTD. The mock samples also have a large number of SNe in order to
minimise the statistical error and isolate the systematic effect
due to the binning. We then rebin the SFHs of the sample into the same
three coarse time bins we have used for the LOSS-SDSS analysis, 
and we recover the DTD binned into those same three time 
intervals. Finally, we compare the integrals $N_{\rm SN}/M$ 
over the input DTD and the recovered DTD.
We find that the $N_{\rm SN}/M$ of the recovered DTD systematically 
underestimates the input $N_{\rm SN}/M$ by 10--20\%, with the larger values
for input DTDs that rise more steeply at small delays. This systematic
error is 
comparable to the statistical errors in $N_{\rm SN}/M$ we have found
for the LOSS-DTD sample. Larger SN samples will naturally allow finer
temporal binning of the DTD, reducing this systematic effect.
   
\section{Conclusions and Outlook}

We have presented a method to recover the SN DTD by jointly analysing 
SN survey data and reconstructed SFHs of the individual galaxies in
a survey, while
accounting for the small-number, Poisson-statistical nature of SN
events in those individual galaxies. 
Our method is an improvement over previous ones in that 
it uses the full information
contained in the data and avoids unnecessary averaging. 
The method is based on casting
the expected SN numbers in each galaxy in the survey as a set of
linear equations, in which the parameters to be constrained by the data
are the values of the DTD in binned time intervals. We have shown 
a maximum-likelihood method 
by which to invert those equations to recover the DTD,
and demonstrated the method's performance using simulated mock samples.
We have applied the method to a sample consisting of those
galaxies in the LOSS SN survey that have SDSS-VESPA SFH
reconstructions, along with the SNe that these galaxies have hosted during
the LOSS. 

The recovered DTDs for  CC~SNe and SNe~Ia are limited by Poisson
statistics, because of the small numbers of SNe in this restricted
subset of LOSS, and by systematic errors, because for many galaxies
the light within the SDSS fibre aperture does not reliably represent
the full stellar population of the galaxy and its SFH.
Despite these shortcomings, we are able to make the following
statements based on analysis of these data.\\
(1) A ``prompt'' SN~Ia population, defined here as one that explodes within
420~Myr of star formation, exists at $>99$\% confidence.
This confirms the growing number of reports of such a population, at
levels similar to those found here. 
\\
(2) We clearly detect, at the $4\sigma$ level, a ``delayed'' SN~Ia population 
  with delays in the range 2.4--13~Gyr, and a mean rate over this
  interval of $\Psi_3=(2.6\pm 0.7) \times 10^{-4}$ SNe yr$^{-1}$
per $10^{10}\, {\rm M}_\odot$ formed, or
$A=(5.1\pm 1.4) \times 10^{-4}$ SNe yr$^{-1}$
per remaining $10^{10}\, {\rm M}_\odot$ in an old population.
These first two conclusions together indicate that the SN~Ia DTD
peaks at short delays, but extends over
 a broad range of delay times, out to at least several Gyr.
Progenitor models, a few of which we have briefly compared to 
our recovered DTD, will need to reproduce the observed numbers. \\
(3) The time-integrated SN~Ia yield is 
$N_{\rm Ia}/M=(2.3\pm 0.6)\times 10^{-3}$  
SNe per unit solar mass formed, or $(4.6\pm 1.2)\times
10^{-3}$ SNe~Ia per remaining unit solar mass in an old population.
The best-fit value is a factor of 1.5--3 lower than the corresponding
number in galaxy clusters, as deduced from their measured
iron-to-stellar mass ratios. Although the current uncertainties
 can still accommodate this difference, it may indicate that clusters underwent
additional enrichment by CC~SNe from an early stellar population with
a top-heavy IMF, or that SN~Ia production is more efficient in 
galaxy clusters than in the field. The latter possibility has support
 both from direct cluster SN rate measurements and from cluster
 element abundance analysis.   The measured 
time-integrated SN~Ia yield also implies 
$\eta\approx 8\%$ for the exploding fraction among the 
parent population of SN~Ia primaries,
 if assumed to come from the 3--8 M$_\odot$ initial mass range. \\
(4) The time-integrated CC~SN yield is 
$N_{\rm CC}/M=(1.0\pm 0.2)\times 10^{-2}$  
SNe per unit solar mass formed. This rules out low-mass limits for
CC explosions that are much above 8~M$_\odot$. Conversely, scenarios in which 
a significant fraction of high-mass stars end their evolution without
SN explosions are excluded, unless the low-mass limit for core
collapse is significantly below 8 M$_\odot$.\\
(5) The ratio of CC~SN to SN~Ia numbers from a brief
burst of star formation, integrated over a Hubble time,
 is $(4^{+3}_{-1.5})$. 

Our work points to the kinds of
data that would improve upon these results. First, a larger sample
of survey galaxies with spectroscopy, and hence SFHs, would obviously
reduce the Poisson errors and permit better temporal resolution. As there
are 14882 galaxies monitored by LOSS, obtaining spectra for most or
all of them would be a large, but not impossible, task. A dedicated
or partly dedicated 2--4~m-class telescope could achieve this on a 
few-year timescale. Long-slit spectra, drifted across the galaxy
perpendicular to the slit length, would be more representative of the
entire stellar population of each galaxy, largely avoiding the
small-fibre-aperture problem we have encountered.
 Ideally, instead of long-slit spectra
of the LOSS galaxies, one would obtain integral-field spectroscopy
of each of these galaxies (using, e.g., an instrument such as SAURON
on the William Herschel 4.2~m telescope;
Bacon et al. 2001). 
In addition to including the full
SFH of each galaxy without any aperture losses, such data, in the
context of our method, would easily allow breaking up each galaxy into
independent subunits, and considering the SFH of each subunit in
relation to the SNe that it hosted (or, for almost all subunits, the SNe that
it did not host). 

It might be objected at this point, that 
because of the random stellar velocities in
each galaxy, SN
progenitor stars diffuse away from the regions where they were formed, and
that this would invalidate our approach; 
by the time a SN~Ia exploded, it would
reside within a region that is completely different from
the one in which its progenitor was formed. 
While this is
true, and it would in fact invalidate a traditional ``SN delay time''
analysis in which a single characteristic stellar population 
age is assigned to a region,
it is inconsequential to our current method, in which the entire SFH
of the region is considered. The reason is that the same
spatial diffusion affects both the progenitor population and those
stars among it that eventually explode. 
To see this, consider, as a toy example, a
grid of $3\times 3$ adjacent ``cells'' in a particular galaxy. 
Suppose, for example, that 500~Myr
ago there was a short burst of star formation in the central cell,
forming a stellar mass $M$, and no
activity in the other cells. Suppose, further, that the SN DTD is such that
the stellar population formed in the burst leads, 500~Myr
later, to nine SNe over the course of a decade, 
which are therefore detected in this galaxy by a SN survey such as
LOSS (this is admittedly a somewhat unrealistically large number of
SNe, but is used here just for the sake of illustration).
The galaxy has thus produced 
a ratio of $9/M$ SNe per unit stellar mass formed 500~Myr ago. 
Finally, suppose that the stellar diffusion timescale in the galaxy
is such that, over the 500~Myr,
 the progenitors of the 9 SNe, before exploding, have drifted out of
 the central cell in which they
were formed, and there is now, on average, one SN in each
cell. However, the entire stellar population of the burst will have
diffused in the same way, and therefore each cell will have 1/9 of the
500-Myr-old population that was originally in the central cell. When
we compare SN numbers to the 500~Myr-old stellar mass present today
in each cell, we will see 1 SN per $M/9$ of stellar mass formed. In the
DTD we derive, we will therefore still deduce the correct ratio of $9/M $ 
SNe per unit stellar mass
formed 500~Myr earlier. 

This argument holds, no matter what are the lookback times or the
diffusion timescales. It also holds for arbitrarily complex SFHs,
which can be viewed as linear superpositions in space and in time 
of toy models of the type above. 
If a past starburst at a certain place
and time in a galaxy produces SNe that are observed in the course of a
survey, the stellar population of the burst and the SNe that it produces
will both diffuse in the same way. If, on the other hand, a specific burst
does not produce SNe detected by the survey (because the DTD has a low
amplitude at the corresponding delay), then there will be no
correlation between the number of SNe per cell and the mass of stars 
of that age per cell.
An individual cell hosting SNe
may, of course, include unrelated stellar populations that
did not produce those SNe, but whose stars nonetheless drifted into the
cell. However, over the entire galaxy, there will be no correlation
between SNe and stars of that particular age, and it is such
correlations that drive the results of our DTD recovery method.

We note that for the integral-field spectroscopy approach to work, 
the signal-to-noise ratio of
the spectra of each individual galaxy cell needs to be sufficiently high 
for a reliable SFH reconstruction 
to be obtained. In particular, the presence 
of old stellar populations that are superimposed on younger and more
luminous stars must be detectable. 
Naturally, spatially resolved, medium-spectral-resolution data
 of such a large sample of nearby galaxies would find many additional 
applications, and hence such data are worth the large effort required.

Shortly before submission of this work, Brandt et al. (2010) presented
a DTD reconstruction analysis of a different SN sample. Their
methodology shares several elements with ours. Brandt et al. (2010) 
study 107 SNe~Ia from SDSS-II.
Like us, they use VESPA to derive SFHs for a sample of SDSS galaxies,
binned into three time bins, identical to those we have chosen. Like
us, they treat the DTD amplitudes in the three discrete bins as
free parameters, which are determined by a maximum-likelihood
procedure. 
However, rather than
comparing directly the presence or absence of SNe in each galaxy to 
the predictions of the DTD model (as we have), they use the DTD to 
create mock SN-host samples, and compare the mean spectrum of
the mock host samples to the mean spectrum of the real host galaxies. 
Brandt et al. (2010) reach similar conclusions
 to ours, namely, significant detections of both prompt ($<420$~Myr)
and delayed ($>2.4$~Gyr) SN~Ia DTD components.

\section*{Acknowledgments}
We thank Rita Tojeiro for her assistance and patience with obtaining 
and explaining the VESPA SFHs, Jesse Leaman for his contribution
to the determination of LOSS SN rates, and Keren Sharon for
her help with the comparison to models. The anonymous referee is
thanked for constructive comments. D.M acknowledges support 
by a grant from the Israel Science Foundation. LOSS, conducted
by A.V.F.'s group at the University of California, Berkeley, 
has been supported by many grants from the US National
Science Foundation (most recently AST-0607485 and AST-0908886),
the TABASGO Foundation, US Department of Energy SciDAC grant
DE-FC02-06ER41453, and US Department of Energy grant
DE-FG02-08ER41653. KAIT and its ongoing operation were made possible
by donations from Sun Microsystems, Inc., the Hewlett-Packard Company,
AutoScope Corporation, Lick Observatory, the National Science Foundation, 
the University of
California, the Sylvia \& Jim Katzman Foundation, and the TABASGO
Foundation.



\begin{thebibliography}{}

\bibitem[Aubourg et al.(2007)]{2007arXiv0707.1328A} Aubourg, E.,
  Tojeiro, R., Jimenez, R., Heavens, A.~F., Strauss, M.~A., \&
  Spergel, D.~N.\ 2008, \aap, 492, 631
\bibitem[Bacon et al.(2001)]{2001MNRAS.326...23B} Bacon, R., et al.\ 2001, 
\mnras, 326, 23 
\bibitem[Bell et al.(2003)]	{2003ApJS..149..289B} Bell, E.~F., McIntosh, D.~H., Katz, N., \& Weinberg, M.~D.\ 2003, \apjs, 149, 289
\bibitem[]{2009ARep...53..214B}Bogomazov, A.I \& Tutukov, A.V., 2009, Astronomy Reports,
  53, 214
\bibitem[]{}Brandt, T.~D., Tojeiro, R., Aubourg, E., Heavens, A.,
  Jimenez, R., \& Strauss, M.~A.\ 2010, arXiv:1002.0848
\bibitem[Bruzual 
\& Charlot(2003)]{2003MNRAS.344.1000B} Bruzual, G., \& Charlot, S.\ 2003, \mnras, 344, 1000 
\bibitem[Dahlen et al.(2004)]	{2004ApJ...613..189D} Dahlen, T., et al.\ 2004, \apj, 613, 189 
\bibitem[Dahlen et al.(2008)]{2008ApJ...681..462D} Dahlen, T., Strolger, L.-G., \& Riess, A.~G.\ 2008, \apj, 681, 462 
\bibitem[de Plaa et al.(2007)]{2007A&A...465..345D} de Plaa, J., Werner, N., Bleeker, J.~A.~M., Vink, J., Kaastra, J.~S., \& M{\'e}ndez, M.\ 2007, \aap, 465, 345
\bibitem[Della Valle et al.(2005)]{2005ApJ...629..750D} Della Valle, M., Panagia, N., Padovani, P., Cappellaro, E., Mannucci, F., \& Turatto, M.\ 2005, \apj, 629, 750
\bibitem[Filippenko (1997)]{} Filippenko, A.~V. 1997, ARAA, 35, 309
\bibitem[Filippenko et al.(2001)]{2001ASPC..246..121F} Filippenko, A.~V.,
Li, W., Treffers, R.~R., \& Modjaz, M.\ 2001, in Small-Telescope 
Astronomy on Global Scales, ed. W. P. Chen, C. Lemme, \& 
B. Paczy\'{n}ski (San Francisco: ASP), 121
\bibitem[]{}Filippenko, A.V., et al. 2010, in preparation
\bibitem[Gal-Yam \& Maoz(2004)]	{2004MNRAS.347..942G} Gal-Yam, A., \&
  Maoz, D.\ 2004, \mnras, 347, 942
\bibitem[Gonz{\'a}lez Hern{\'a}ndez et al.(2009)]{2009ApJ...691....1G} 
Gonz{\'a}lez Hern{\'a}ndez, J.~I., Ruiz-Lapuente, P., Filippenko, A.~V., 
Foley, R.~J., Gal-Yam, A., \& Simon, J.~D.\ 2009, \apj, 691, 1 
\bibitem[Graham et al.(2008)]{2008AJ....135.1343G} Graham, M.~L., et al.\ 
2008, \aj, 135, 1343 
\bibitem[Greggio(2005)]		{2005A&A...441.1055G} Greggio, L.\ 2005, \aap, 441, 1055
\bibitem[]{}Greggio, L., Renzini, A., \& Daddi, E. 2008, MNRAS, 388, 829 
\bibitem[Han 
\& Podsiadlowski(2004)]{2004MNRAS.350.1301H} Han, Z., \& Podsiadlowski, P.\ 2004, \mnras, 350, 1301 
\bibitem[Hurley et al.(2002)]{2002MNRAS.329..897H} Hurley, J.~R., Tout, 
C.~A., \& Pols, O.~R.\ 2002, \mnras, 329, 897 
\bibitem[Iben \& Tutukov(1984)]{1984ApJS...54..335I} Iben, I., Jr., \& 
Tutukov, A.~V.\ 1984, \apjs, 54, 335
\bibitem[Kerzendorf et al.(2009)]{2009ApJ...701.1665K} Kerzendorf, W.~E., 
Schmidt, B.~P., Asplund, M., Nomoto, K., Podsiadlowski, P., Frebel, A., 
Fesen, R.~A., \& Yong, D.\ 2009, \apj, 701, 1665 
\bibitem[Kobayashi et al.(2000)]{2000ApJ...539...26K} Kobayashi, C., 
Tsujimoto, T., \& Nomoto, K.\ 2000, \apj, 539, 26
 \bibitem[Kochanek et al.(2008)]{2008ApJ...684.1336K} Kochanek, C.~S., 
Beacom, J.~F., Kistler, M.~D., Prieto, J.~L., Stanek, K.~Z., Thompson, 
T.~A., \& Yuumlksel, H.\ 2008, \apj, 684, 1336 
\bibitem[Kroupa(2007)]{2007astro.ph..3124K} Kroupa, P.\ 2007, 
arXiv:astro-ph/0703124 
\bibitem[]{}Leaman, J., et al. 2010, in preparation
\bibitem[Li et al.(2000)]{2000AIPC..522..103L} Li, W., et al.\ 2000,
American Institute of Physics Conference Series, 522, 103
\bibitem[]{}Li, W., et al. 2010a, in preparation
\bibitem[]{}Li, W., et al. 2010b, in preparation
\bibitem[{{Mannucci} {et~al.}(2005){Mannucci}, {Della Valle}, {Panagia},
  {Cappellaro}, {Cresci}, {Maiolino}, {Petrosian}, \& {Turatto}}]{Mannucci_05}
{Mannucci}, F., {Della Valle}, M., {Panagia}, N., {Cappellaro}, E., {Cresci},
  G., {Maiolino}, R., {Petrosian}, A., \& {Turatto}, M. 2005, \aap,
  433, 807
\bibitem[{{Mannucci} {et~al.}(2006){Mannucci}, {Della Valle}, \&
  {Panagia}}]{Mannucci_06}
{Mannucci}, F., {Della Valle}, M., \& {Panagia}, N. 2006, \mnras, 370, 773
\bibitem[Mannucci et al.(2008)]{2008MNRAS.383.1121M} Mannucci, F., Maoz, 
D., Sharon, K., Botticella, M.~T., Della Valle, M., Gal-Yam, A., 
\& Panagia, N.\ 2008, \mnras, 383, 1121 
\bibitem[Maoz \& Gal-Yam(2004)]	{2004MNRAS.347..951M} Maoz, D., \& Gal-Yam, A.\ 2004, \mnras, 347, 951
\bibitem[Maoz(2008)]{2008MNRAS.384..267M} Maoz, D.\ 2008, \mnras, 384, 267 
\bibitem[Maoz \& Mannucci(2008)]{2008MNRAS.388..421M} Maoz, D., \& Mannucci, F.\ 2008, \mnras, 388, 421
\bibitem[]{}Maoz, D., \& Badenes, C.. 2010, MNRAS, submitted, arXiv:1003.3031
\bibitem[]{}Maoz, D., et al. 2010, in preparation
\bibitem[Maraston(2005)]{2005MNRAS.362..799M} Maraston, C.\ 2005, \mnras, 
362, 799 
\bibitem[]{}Mennekens, N., Vanbeveren, D., De Greve, J.~P., De Donder,
  E., \aap, in press, arXiv:1003.2491
\bibitem[]{}Meng, X., \& Yang, W.\ 2010, \apj, 710, 1310
\bibitem[Nelemans et al.(2005)]{2005AA...440.1087N} Nelemans, G., et al.\ 
2005, \aap, 440, 1087 
\bibitem[Pritchet et al. (2008)]{2008ApJ...683L..25P} Pritchet, C.~J.,
  Howell, D.~A., \& Sullivan, M.\ 2008, \apjl, 683, L25
\bibitem[Roelofs et al.(2008)]{2008MNRAS.391..290N} Roelofs, G.,
  Bassa, C., Voss, R.,  \& Nelemans, G.\ 2008, \mnras, 391, 290
\bibitem[Poznanski et al.(2007)]{2007MNRAS.382.1169P} Poznanski, D., et 
al.\ 2007, \mnras, 382, 1169 
\bibitem[Press et al.(1992)]{1992nrfa.book.....P} Press, W.~H., Teukolsky, 
S.~A., Vetterling, W.~T., 
\& Flannery, B.~P.\ 1992, Cambridge: University Press, |c1992, 2nd ed.,  
\bibitem[Raskin et al.(2008)]{2008ApJ...689..358R} Raskin, C., Scannapieco,
E., Rhoads, J., \& Della Valle, M.\ 2008, \apj, 689, 358
\bibitem[Raskin et al.(2009)]{2009MNRAS.399L.156R} Raskin, C., Timmes, 
F.~X., Scannapieco, E., Diehl, S., \& Fryer, C.\ 2009a, \mnras, 399, L156 
\bibitem[Raskin et al.(2009)]{2009ApJ...707...74R} Raskin, C., Scannapieco,
E., Rhoads, J., \& Della Valle, M.\ 2009b, \apj, 707, 74
\bibitem[Rosswog et al.(2009)]{2009ApJ...705L.128R} Rosswog, S., Kasen, D., 
Guillochon, J., \& Ramirez-Ruiz, E.\ 2009, \apjl, 705, L128 
\bibitem[Ruiter et al.(2009)]{2009ApJ...699.2026R} Ruiter, A.~J., 
Belczynski, K., \& Fryer, C.\ 2009, \apj, 699, 2026 
\bibitem[Salpeter(1955)]{1955ApJ...121..161S} Salpeter, E.~E.\ 1955, \apj, 121, 161 
 \bibitem[Scannapieco \& Bildsten(2005)]{2005ApJ...629L..85S} Scannapieco, E., \& Bildsten, L.\ 2005, \apjl, 629, L85
\bibitem[Sharon et al.(2007)]{2007ApJ...660.1165S} Sharon, K., Gal-Yam, A., Maoz, D., Filippenko, A.~V., \& Guhathakurta, P.\ 2007, \apj, 660, 1165 
\bibitem[Smartt(2009)]{2009ARA&A..47...63S} Smartt, S.~J.\ 2009,
  \araa, 47, 63
\bibitem[Strolger et al.(2004)]{2004ApJ...613..200S} Strolger, L.-G., et 
al.\ 2004, \apj, 613, 200 
\bibitem[Sullivan et al.(2006b)]{2006ApJ...648..868S} Sullivan, M., et al. 2006, \apj, 648, 868
\bibitem[Tojeiro et al.(2007)]{2007MNRAS.381.1252T} Tojeiro, R., Heavens, 
A.~F., Jimenez, R., \& Panter, B.\ 2007, \mnras, 381, 1252 
\bibitem[Tojeiro et al.(2009)]{2009ApJS..185....1T} Tojeiro, R., Wilkins, 
S., Heavens, A.~F., Panter, B., \& Jimenez, R.\ 2009, \apjs, 185, 1 
\bibitem[Totani et al.(2008)]{2008arXiv0804.0909T} Totani, T.,
  Morokuma, T., Oda, T., Doi, M., \& Yasuda, N.\ 2008, \pasj, 60, 1327
\bibitem[Tout(2005)]{2005ASPC..330..279T} Tout, C.~A.\ 2005, in The 
Astrophysics of Cataclysmic Variables and Related Objects, 330, 279 
\bibitem[Tutukov \& Yungelson(2002)]{} Tutukov, A.~V., \& Yungelson,
  L.~R.\ 2002, Astronomy Reports, 46, 667
\bibitem[Voss 
\& Nelemans(2008)]{2008Natur.451..802V} Voss, R., \& Nelemans, G.\ 2008, \nat, 451, 802 
\bibitem[Webbink(1984)]{1984ApJ...277..355W} Webbink, R.~F.\ 1984, \apj, 
277, 355 
\bibitem[Whelan \& Iben(1973)]{1973ApJ...186.1007W} Whelan, J., \& Iben, 
I.~J.\ 1973, \apj, 186, 1007 
\bibitem[York et al.(2000)]{2000AJ....120.1579Y} York, D.~G., et al.\ 2000, 
\aj, 120, 1579 

\end{thebibliography}
\end{document}